\documentclass[manuscript, screen]{acmart}
\usepackage{booktabs} 

\usepackage[ruled]{algorithm2e} 

\usepackage{colortbl}
\usepackage{xcolor}
\usepackage{multicol}
\usepackage{multirow}
\usepackage{longtable}
\usepackage{booktabs}

\usepackage{footnote}
\makesavenoteenv{tabular}
\usepackage{savefnmark}

\newcommand{\classificationschema}{
\small
	\begin{tabular}{@{}p{1.6cm}p{13cm}@{}}
		\toprule
			\underline{\textbf{Action}} & Time intervals denoting interactions with resources, further breaking down task switches. \\[1ex]
			\textbf{Interaction} & When users provide input to a resource through either the keyboard (e.g., writing) or the mouse (e.g., selecting text, navigating menus to access operations like saving). \\
			\textbf{Navigation within} & When users navigate within a resource while the majority of the content of the resource is already visible (e.g., the use of arrow keys to move the cursor within a text; scrolling using either the mouse wheel or scrollbar; moving, minimizing, or resizing a window; reading; concentrated eye movements within the active task). This does not require an application window to have focus (i.e., be active). The same resource remains visible during the annotated interval. As opposed to the `Interaction' category, the content (or selection) of the resource does not change. \\
			\textbf{Navigation to opened} & When users restore an open resource, while the resource is not yet visible. This action can be directed at fully hidden resources (e.g., accessing the taskbar, using Alt-Tab) as well as partially hidden resources in the background (e.g., bringing windows covered by others to the foreground). As opposed to the `Navigation within' category, the majority of the content of the resource is hidden. The resource does not need to end up visible at the end of the annotated interval (for example, when using a preview). One interesting scenario is using the taskbar preview function of Windows to thoroughly inspect a resource for a long period of time, this is annotated as `Navigation to opened', followed by `Navigation within'. \\
			\textbf{Navigation to closed} & When users open a closed resource. Most commonly, when no application window for the resource is open. However, resources are defined as containers of content which can be shown or hidden separately, thus reusing, or opening new tabs is also considered as navigation to a closed resource. Likewise is navigating to a different file path using the file explorer. When a resource is already open, but the user's strategy follows that of opening a closed resource, the interval is still labeled as `Navigation to closed'.\\
			\textbf{Pause} & When users interrupt their current action, e.g., by repositioning themselves, by verbally expressing relief, or by closing their eyes for an extended duration.  \\
		\midrule
			\underline{\textbf{Intent}} & Applied to actions to capture their purpose.\\[1ex]
			\textbf{Identify} & When the purpose of the action is the identification of the task description or task resources, rather than making progress on it. E.g., figuring out which resource of the task set is required. This is generally observed at the start of task resumption. \\
			\textbf{Reorganize} & When the purpose of the action is to reorganize resources in the task set. This intent can be observed both during the disengagement stage (e.g., leaving visual clues to facilitate future resumption) as well as during the resumption stage (e.g., restoring required resources to resume work). \\
			\textbf{Task} & When the purpose of the action is to make progress on the current task. When the user reorganizes the workspace while simultaneously working on the task, the `Task' category takes precedence over `Reorganize'.\\
		\midrule
			\underline{\textbf{Interface}} & Applied to actions to record the user interface used. \\[1ex]
			\textbf{Taskbar} & Using the taskbar (including preview functionality) to navigate to a resource. \\
			\textbf{Alt-Tab} & Using the Alt-Tab shortcut key to navigate to a resource. \\
			\textbf{Foreground} & Navigate to a window which is partially or fully visible in the foreground. \\
		\midrule
			\underline{\textbf{Modifier}} & Applied to actions to annotate interesting fragments. \\[1ex]
			\textbf{Error} & Used to highlight an erroneous action given the intent, observable from the data (e.g., the resumption of a wrong resource, the unjustified use of a shortcut).\\
		\bottomrule
	\end{tabular}	
}

\newcommand{\errors}{
	\small
	\begin{tabular}{@{}ccccccc@{}}
		\noalign{\smallskip}
		 									&	\cellcolor[gray]{0.8}\textbf{P1}	
											&	\cellcolor[gray]{0.8}\textbf{P2}	
											&	\cellcolor[gray]{0.8}\textbf{P3}	
											&	\cellcolor[gray]{0.8}\textbf{P4}	
											&	\cellcolor[gray]{0.8}\textbf{P5}	
											&	\cellcolor[gray]{0.8}\textbf{P6}	\\
				\cellcolor[gray]{0.8}\textbf{\# Errors}		&	11 	&	2	&	3	&	4	&	20	&	3		\\
				\cellcolor[gray]{0.8}\textbf{Time}			&	36.1 s	&	7.8 s	&	11.1 s	&	23.2 s	&	85.0 s	&	6.1 s	\\
				\cellcolor[gray]{0.8}\textbf{Percentage}			&	18.2\%	&	3.5\%	&	6.2\%	&	15.2\%	&	15.6\%	&	2.8\%	\\
	\end{tabular}
}

\newcommand{\breakdowns}{
	\small
	\begin{tabular}{@{}cccccc@{}}
		\noalign{\smallskip}
		 									&	\cellcolor[gray]{0.8}\textbf{Interaction}	
											&	\cellcolor[gray]{0.8}\textbf{Within}	
											&	\cellcolor[gray]{0.8}\textbf{Opened}	
											&	\cellcolor[gray]{0.8}\textbf{Closed}	
											&	\cellcolor[gray]{0.8}\textbf{Pause} \\
				\cellcolor[gray]{0.8}\textbf{$\Box$}			&	30.2\% 	&	21.4\%	&	1.8\% (1$^i$)	&	0\%	&	46.6\%		\\
				\cellcolor[gray]{0.8}\textbf{$\triangleright$}	&	0.9\% (3$^i$)	&	39.0\%	&	50.9\%	&	6.5\%	&	2.7\%	\\
	\end{tabular}
}

\newcommand{\winfeatures}{
	\small
	\begin{tabular}{@{}ccccccc@{}}
		\noalign{\smallskip}
		 									&	\cellcolor[gray]{0.8}\textbf{P1}	
											&	\cellcolor[gray]{0.8}\textbf{P2}	
											&	\cellcolor[gray]{0.8}\textbf{P3}	
											&	\cellcolor[gray]{0.8}\textbf{P4}	
											&	\cellcolor[gray]{0.8}\textbf{P5}	
											&	\cellcolor[gray]{0.8}\textbf{P6}	\\
				\cellcolor[gray]{0.8}\textbf{Alt-Tab}		&	12.9\%	&	   0\%	&	15.7\%	&	   0\%	&	   0\%		&	  0\%	\\
				\cellcolor[gray]{0.8}\textbf{Taskbar}		&	83.7\%	&	98.6\%	&	84.3\%	&	84.6\%	&	87.1\%	&	87.2\%	\\
				\cellcolor[gray]{0.8}\textbf{Foreground}	&	 3.4\%	&	 1.4\%	&	   0\%	&	15.4\%	&	12.9\%	&	12.8\%	\\
	\end{tabular}
}

\newcommand{\costPerP}{
	\small
	\begin{tabular}{@{}ccccccc@{}}
		\noalign{\smallskip}
		 									&	\cellcolor[gray]{0.8}\textbf{P1}	
											&	\cellcolor[gray]{0.8}\textbf{P2}	
											&	\cellcolor[gray]{0.8}\textbf{P3}	
											&	\cellcolor[gray]{0.8}\textbf{P4}	
											&	\cellcolor[gray]{0.8}\textbf{P5}	
											&	\cellcolor[gray]{0.8}\textbf{P6}	\\
				\cellcolor[gray]{0.8}\textbf{Average}					&	19.0 s		&	18.6 s		&	14.5 s		&	23.6 s		&	\cellcolor[gray]{0.95}47.6 s		&	19.6 s		\\
				\cellcolor[gray]{0.8}\\[-5pt]
				\cellcolor[gray]{0.8}\textbf{$\Omega^{[1]}$}		&	9:44		&	9:04		&	7:04		&	11:29		&	\cellcolor[gray]{0.95}23:12		&	9:32		\\
				\cellcolor[gray]{0.8}\textbf{$\mathcal{O}^{[2]}$}	&	20:47		&	19:21		&	15:05		&	24:32		&	\cellcolor[gray]{0.95}49:33		&	20:22		\\
	\end{tabular}
}

\usepackage{nicefrac}
\usepackage{lscape}
\usepackage{float}
\usepackage{nameref}

\usepackage[figuresright]{rotating}
\usepackage{rotating}

\begin{document}
\title{The Hidden Cost of Window Management}

\author{Steven Jeuris}
\orcid{0000-0001-9645-8275}
\affiliation{%
  \institution{Technical University of Denmark}
  \streetaddress{Richard Petersens Plads, Building 324, room 270}
  \city{Kgs. Lyngby}
  \postcode{2800}
  \country{Denmark}}
\email{sjeu@dtu.dk}

\author{Paolo Tell}
\affiliation{%
  \institution{IT University of Copenhagen}
  \streetaddress{Rued Langgaards Vej 7, 4D11}
  \city{Copenhagen S}
  \postcode{2300}
  \country{Denmark}
}
\email{pate@itu.dk}

\author{Steven Houben}
\affiliation{%
  \institution{Lancaster University}
  \streetaddress{South Drive, Campus of Lancaster University, Office B19, Infolab}
  \city{Copenhagen S}
  \postcode{LA1 4WA}
  \country{United Kingdom}
}
\email{s.houben@lancaster.ac.uk}

\author{Jakob E. Bardram}
\orcid{0000-0001-9645-8275}
\affiliation{%
  \institution{Technical University of Denmark}
  \streetaddress{Richard Petersens Plads, Building 324, room 160}
  \city{Kgs. Lyngby}
  \postcode{2800}
  \country{Denmark}}
\email{jakba@dtu.dk}

\renewcommand\shortauthors{Jeuris, S. et al}

\begin{abstract}
Most window management systems support multitasking by allowing users to open, resize, position, and switch between application windows.
Although multitasking has become a way of life for most knowledge workers, our current understanding of how users use window management features to switch between multiple tasks---which may comprise multiple application windows---is limited.
In this paper, we present a study providing an in-depth analysis of how task switching is supported in Windows 7.
As part of analysis, we developed an interface-agnostic classification of common task switching operations supported by window managers which can be used to quantify the time spent on each constituting action.
Our study shows that task switching is a time intensive activity and highlights the dominant actions that contribute to task switch time.
Furthermore, our classification highlights the specific operations that are optimized by more recent and experimental window managers and allows identifying opportunities for design that could further reduce the overhead of switching between tasks.
\end{abstract}

%
%


%
%


\maketitle

\section{Introduction}
\label{sec:introduction}

Over the years, the desktop metaphor (which was introduced to allow users to more intuitively interact with computer systems) has been refined to allow for increasingly complex knowledge work, requiring access to a myriad of information resources and frequent switching between different parallel tasks~\cite{gonzalez2004constant}---i.e., {multitasking}. This often leads to {information overload}~\cite{mulder2006information}. Therefore, not surprisingly, current window managers have evolved considerably since their original inception to allow having more and more windows open simultaneously and to provide different mechanisms of switching between them. For example, Windows 10 recently introduced a `task view' supporting virtual desktops, a feature known earlier on OS X as `spaces' and originally introduced as part of research at Xerox PARC~\cite{henderson1986rooms}.

Prior studies have investigated the nature of and the reasons for multitasking. For example, studies have observed knowledge workers within their workspace. This has lead to insights into how often multitasking takes place, the amount of documents used per task, and the different types of interruptions which cause users to switch between tasks~\cite{bannon1983activity, czerwinski2004diary, gonzalez2004constant, hardy2012voluntary, jin2009selfinterruption}.
However, very few studies have analyzed the \emph{process} (entailing \emph{`micro operations'}) of switching between tasks when using a traditional window manager or the strategies adopted by users to this end.
This is an important oversight. As argued by ~\citeN{mulder2006information}: ``\emph{the heart of what information overload really is, may very well lie \emph{between} tasks rather than \emph{within}.}''
This is confirmed by a recent experimental study which showed that switching between tasks when using a window manager supporting virtual desktops (as opposed to no virtual desktops) reduces cognitive load during multitasking~\cite{jeuris2016dedicated}.
As such, there is a need for a more thorough understanding of how users utilize window managers when switching between tasks to better evaluate the impact new features might have and to identify further opportunities for design.

This paper presents a study investigating the window manager of Windows 7 (supporting common window operations still in use today) by unraveling the processes and strategies used when switching from one task to another. The paper introduces three contributions.
First, based on the most detailed study of task switches to date, we derive an interface-agnostic \emph{classification of window management operations}, i.e., a detailed coding framework that provides a structure for analyzing and understanding the procedure of switching from one task to another.
Second, demonstrating how this classification can be used quantitatively, we calculate the cost of task switching in Windows 7 for six representative knowledge workers in terms of time spent on reorganizing the work environment during our experiment.
Third, using the classification as the basis for analysis, we discuss existing window management techniques which automate the trivial and tedious work of reorganizing application windows, and identify further opportunities for design to help users in reorienting themselves within documents and resources when resuming a task.
\section{Related Work}
\label{sec:related_work}

Multitasking has been studied using a variety of methods in multiple fields, which can make it challenging to understand how disparate concepts and results are interrelated~\cite{janssen2015integrating}. We will therefore first define, differentiate, and relate \emph{sequential multitasking} (the focus of this paper) to \emph{concurrent multitasking} and \emph{interruptions} during ongoing work.
Second, we will discuss how window management has been considered in earlier multitasking studies and how this paper contributes to a narrower (largely overlooked) research topic: most related studies report on the \emph{amount} of multitasking that takes place and the \emph{effects} of multitasking on actual knowledge work (e.g., task performance), whereas our focus lies on the \emph{time interval during which task switching takes place} (requiring the retrieval of multiple application windows) and the \emph{mechanics} used to this end on a desktop computer.

\subsection{Sequential Multitasking}
According to \citeN{salvucci2009towards}, \emph{``multitasking can be represented along a continuum in terms of time spent on one task before switching to another''}. On one end of the continuum lies \emph{concurrent multitasking}, or the near simultaneous execution of multiple tasks (a classic example is driving while talking on the phone). As part of `dual-task' studies, it has been shown that this can impact task performance~\cite{strayer2001driven}.
On the other end of the continuum lies \emph{sequential multitasking}, during which users interleave several long-lived tasks that are executed one at a time. Sequential multitasking is part of common everyday knowledge work~\cite{bannon1983activity,czerwinski2004diary,gonzalez2004constant}. In contrast to concurrent multitasking, users switch from one task to another within intervals of several minutes. This process involves retrieving all the necessary resources (e.g., multiple application windows) needed to resume a previously suspended task.

Prior studies that have investigated sequential multitasking usually focus on measuring the effects of interleaving multiple tasks on both task productivity and accuracy. For example, results show an inverted U-relationship between multitasking and productivity; there is thus an optimal amount of task switching that leads to the highest productivity. However, increased levels of multitasking lead to a significant loss in accuracy, indicating a trade-off between productivity and accuracy~\cite{adler2012juggling}. This trade-off is further influenced by task difficulty: easy tasks benefit from multitasking due to the increase in stimulation, but task performance for hard tasks can decrease due to an overload in mental workload~\cite{adler2015effects}. Other studies show that users have a tendency to continue working on more rewarding tasks (with a continuous rate of return) and have a tendency to switch tasks after the completion of subgoals~\cite{duggan2013interleaving,payne2007discretionary}. In addition to subgoal completion, there are several other reasons for switching between tasks~\cite{hardy2012voluntary, jin2009selfinterruption}. Based on flow theory, these can be broadly categorized as either originating from negative (e.g., frustration, exhaustion) or positive (e.g., exploration, reorganization) feelings associated with the task~\cite{adler2013self}.

Task switches are not always self-initiated (due to an internal interruption). \emph{Interruption studies} mainly focus on external interruptions, which are defined as short-lived secondary tasks interrupting a primary task (e.g., answering a question posed by a colleague). 
What sets interruption studies apart from sequential multitasking studies (besides their primary focus on external interruptions) is that interrupting tasks are completed instantaneously, in contrast with tasks within sequential multitasking which are long-lived.
Although there is thus a slight difference in focus, primarily reflected in the research questions brought forward, methods and results do overlap.
As part of interruption studies, the effects on both the primary and secondary task are measured.
For example, studies have shown that the disruptive nature of interruptions reduces task performance~\cite{monk2008effect}, can lead to annoyance and anxiety~\cite{bailey2001effects}, and feelings of stress and frustration~\cite{mark2008stress,mark2012email}.
However, the impact of interruptions depends on when they occur: interruptions at task boundaries cause less anxiety and induce less errors~\cite{bailey2006need}, but this might depend on the relevance of the interruption to the primary task~\cite{gould2013does}. Other findings suggest that resuming a primary task slowly can reduce the amount of errors made~\cite{brumby2013recovering}.
Lastly, interruptions can also disrupt task management, i.e., cause the resumption of unintended tasks~\cite{duggan2013interleaving}.

Although our focus lies on sequential multitasking, some findings related to concurrent multitasking are relevant to the study presented in this paper. During concurrent multitasking, a person's cognitive resources compete to execute a set of parallel tasks. In particular, executive function (including working memory, reasoning, and problem solving), relied on heavily as part of knowledge work, can only pursue one goal at a time.
Accordingly, some studies show that when users are aware about incomplete tasks they need to return to at a later point in time (unfulfilled goals), task performance is reduced when working on a task that requires executive function~\cite{marien2012unconscious,masicampo2011unfulfilled}.
However, such negative effects can be overcome entirely by consciously formulating plans for the unfulfilled goals~\cite{masicampo2011consider}.

\subsection{Window Management}
Window managers provide users with features to structure (resize and position), hide, and retrieve application windows, thus supporting concurrent and sequential multitasking.
In contrast, experimental studies on sequential multitasking (as listed above) usually employ a custom window manager allowing participants to switch between trivial full screen tasks (e.g., solving a Sudoku, finding the ``Odd One Out'') by the press of a button~\cite{adler2012juggling,adler2015effects,payne2007discretionary}. This does not represent switching between tasks in a desktop environment where complex tasks require retrieving several application windows using a window manager.
As an early analysis of window usage by \citeN{henderson1986rooms} revealed, users often move back and forth between different small \emph{sets} of windows.

Field studies that do observe how window managers are used in practice have focused mostly on the different strategies by which users organize application windows and the number of window operations that occur. A recurring result is users who use larger display surfaces (large screens or multiple monitors) do not merely use it as `additional space' but give additional meaning to partitions of it (e.g., a secondary task, different type of work, reminders)~\cite{grudin2001partitioning,hutchings2004revisiting}. Similar results are observed when using virtual desktops~\cite{ringel2003analysis}. When users can leave more windows open simultaneously, they perform less window operations: less moving and resizing~\cite{kang2008lightweight} and less windows switches and mouse clicks~\cite{czerwinski2003toward,hutchings2004displayspace,ling2017twobetter}.
Other studies focus on real-world interruptions (e.g., incoming messages)~\cite{iqbal2007disruption}: their frequency, how much time users spend on handling them (long due to potential loss of context), and influencing factors (visual cues may serve as a reminder to resume tasks sooner). Lastly, an ethnographic study observed how using multiple computers to work on multiple tasks provides better support for task switching, not provided by multiple monitors: easier resumption and a clear cognitive separation of tasks~\cite{beale2007multiple}.

Only a handful of experimental window management studies \emph{compare different work environments} supporting task switching.
Early work looked at the impact of larger versus smaller display surfaces on task performance: both larger screens~\cite{czerwinski2003toward} and multiple monitors~\cite{colvin2004productivity,kang2008lightweight,owens2012examination} are preferred by users and result in higher productivity and less errors made when multitasking. One study reported on a reduced workload~\cite{kang2008lightweight}.
A more recent study~\cite{ling2017twobetter}, comparing single- and dual-monitor setups while in addition adding task complexity as a factor, was unable to observe a difference in task performance. The authors argue this is likely due to the nature of their tasks which were more complex than previous studies (required more thinking), thus introducing higher variance in their measures.
A similar observation was made in a recent study observing no difference in task performance between a traditional Windows 7 environment and virtual desktops during task switching~\cite{jeuris2016dedicated}.
However, in addition to task performance, they measured task resumption time and experienced cognitive load, both of which were reduced when using virtual desktops.
Interestingly, a high variance in task resumption time was observed when using the traditional desktop interface, indicating greatly differing task switching strategies among users.

In summary: compared to prior studies, rather than analyzing the impact of task switching on task performance, we investigate the nature of task switching in sequential multitasking as supported by window managers in great detail.
We aim to understand differences among individual users and to gain new insights into how modern window managers can further improve task switching.
\section{Method}
\label{sec:method}

The goal of this study is to investigate \emph{how window managers support the user in switching between separate tasks during sequential multitasking}.
In addition, we assess the \emph{usability} of a traditional window manager by investigating task switching strategies and the problems users encounter.
Similar to other sequential multitasking studies, participants work on a set of different tasks~\cite{adler2012juggling,adler2015effects}. Since our focus lies on investigating the nature of task switches rather than degree of multitasking, we control for task switches by instructing participants to switch between given tasks at predetermined time intervals (mandatory task interleaving). This mimics a heavy multitasking scenario with concurrent deadlines.

\begin{figure}
\small
\setlength{\unitlength}{1cm}
\centering
\begin{picture}(14,1.5)
\put(0,0.5){\makebox(0,0)[l]{\emph{Time}}}
\put(1.1,0.5){\vector(1,0){12.9}}
\put(3.1,0.3){\line(0,1){1.2}}
\put(1.1,0.5){\makebox(0,1)[l]{Task A}}
\put(5.4,0.5){\makebox(0,1)[c]{Disengagement stage}}
\put(7.7,0.5){\line(0,1){1}}
\put(9.75,0.5){\makebox(0,1)[c]{Resumption stage}}
\put(9.75,0){\makebox(0,0)[c]{\parbox{1.8cm}{\centering\emph{Resumption time}}}}
\put(11.8,0.5){\line(0,1){1}}
\put(13.8,0.5){\makebox(0,1)[r]{Task B}}
\put(7.7,0.3){\line(0,1){1.2}}
\put(11.8,0.3){\line(0,1){1.2}}
\put(5.4,0){\makebox(0,0)[c]{\parbox{1.8cm}{\centering\emph{Disengagement time}}}}
\put(3.1,0){\makebox(0,0)[c]{\emph{Interruption}}}
\put(7.7,0){\makebox(0,0)[c]{\emph{Switch}}}
\put(11.8,0){\makebox(0,0)[c]{\emph{Resumption}}}
\end{picture}
\vspace{0.3cm}
  \caption{A task switch, initiated by an (internal or external) interruption,  comprising a disengagement and resumption stage.}
  \label{fig:firstPage}
\end{figure}
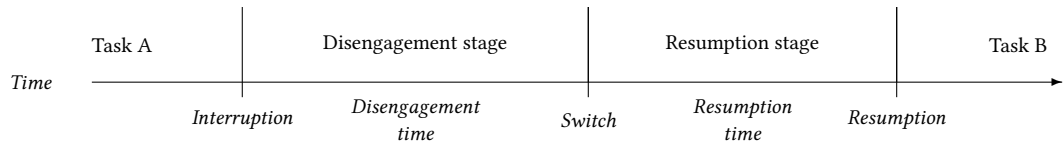

As the starting point of analysis, we adopt the concepts of a \emph{disengagement stage} and a \emph{resumption stage} which make up a task switch, illustrated in Fig.~\ref{fig:firstPage}, building on terminology used in interruption studies~\cite{altmann2004task,boehm2009reducing,trafton2003preparing}.
These time intervals will be further refined and broken down as part of the study presented in this paper.
In contrast to interruption studies, users do not return to a primary task, but rather switch between several long-lived tasks as part of an ongoing multitasking session.
During the disengagement stage, users wrap up work on the previous task and clean up the work environment if they so desire.
During the resumption stage, users prepare the work environment for the next task, which requires the retrieval of all the necessary resources.
We define \emph{task disengagement time} as the time between the initial \emph{interruption} and the first action taken by the user targeted at retrieving a resource belonging to the next task (indicated in Fig.~\ref{fig:firstPage} as `\emph{Switch}').
Although real-world interruptions may be initiated by users themselves (internal interruptions), in the present study all interruptions are external alerts given by the experimenter.
\emph{Task resumption time} is the time between commencing retrieval of the first needed resource (\emph{switch}) and the first work performed on the next task (labeled as `\emph{Resumption}').

\subsection{Participants}
Seven participants (age 22 to 33, 6 male) with varying backgrounds (research, marketing, game design, student) were recruited for the study, each performing 12 task switches over the course of 50 minutes.
Hence, this study covers 84 task switches and 5.8 hours of recorded multitasking.
This was deemed sufficient to reveal several severe usability problems based on our in-depth data analysis.
We selected participants with sufficient, but differing, levels of experience with Microsoft Windows to investigate how experience affects task switching. On a scale from Novice to Expert, no participants selected Novice, three selected Average (P2, P5, Px\footnote{This participant is excluded from data analysis due to a misunderstanding of the given tasks.}), one selected Advanced (P1), and three selected Expert (P3, P4, P6). We consider all participants knowledge workers since they reported using a computer the majority of the day for both work and leisure ($Avg=10.9$ hours, $SD=4.4$ hours). All participants indicated engaging in multitasking on a regular basis: they indicated working on several activities in parallel ($Avg=2.8$ activities, $SD=0.4$ activities). No compensation was given for participating in the study.

\subsection{Materials}
A dedicated Windows 7 notebook with a 15.6 inch screen and external mouse was used for the study.  Participants were allowed to configure the notebook to their liking (e.g., change the keyboard layout and default browser).  Both recently accessed documents and browsing history were cleared prior to each test.  A video camera was set up so that the screen, the participant's hands, and posture were in view.
The screen, webcam (facial expressions), and important key strokes (e.g., Alt-Tab) were recorded using TechSmith Camtasia 8.

\begin{figure}
	\centering
	\includegraphics[width=0.8\columnwidth]{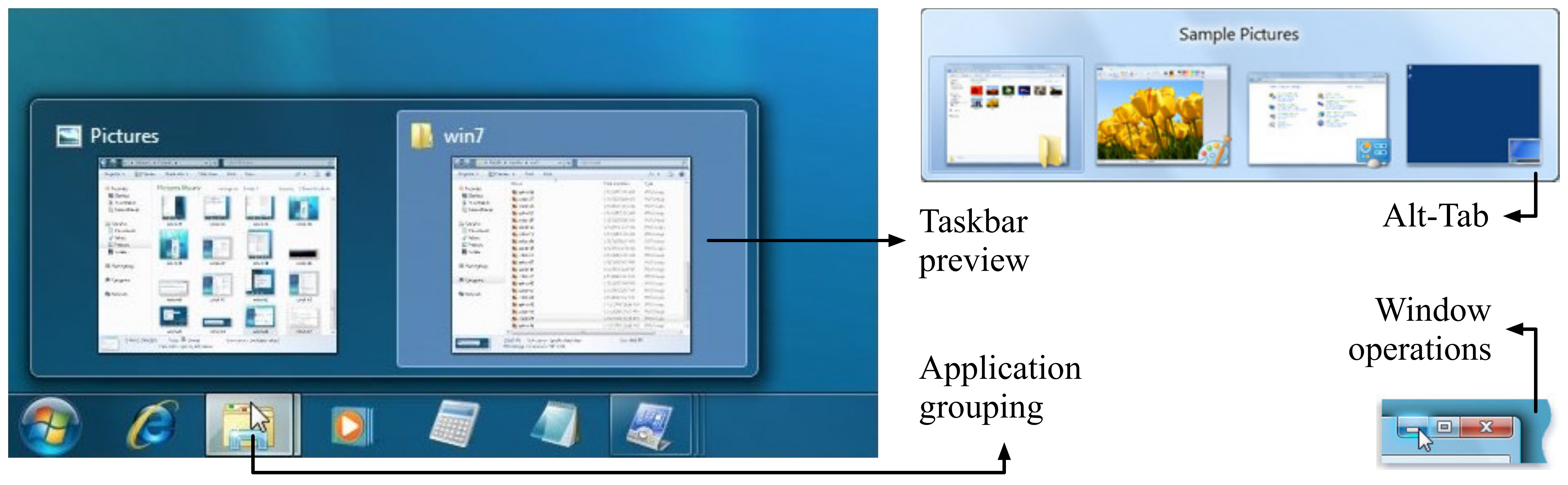}
	\caption{Window manager features of Windows 7 (source: \url{http://windows.microsoft.com}).}
	\label{fig:win7}
\end{figure}

Windows 7 includes a number of features designed to support multitasking (Fig.~\ref{fig:win7}). All open application windows are depicted as icons on a taskbar. Hovering over them shows a window preview. Windows of the same application are grouped together, in which case the preview shows multiple windows. Clicking the taskbar icon, window preview, or the window itself brings it to the foreground (on top of all other windows). Prolonged hovering over thumbnails causes all open windows except the highlighted window to be hidden until the mouse is moved away. Windows can be positioned freely, resized, minimized to the taskbar, maximized full screen, or docked side-by-side.
These functions are also available using shortcut keys. An additional shortcut key hides all windows, thus revealing the desktop. Lastly, pressing Alt-Tab brings up an overview screen which displays all open application windows. 
From here, the user can navigate between windows by repeatedly using Alt-Tab until the desired window is selected. Upon release the selected window is brought to the foreground. Alternatively, the mouse can be used to select a window directly from the Alt-Tab overview.

\begin{figure}
  \centering
  \includegraphics[width=0.85\columnwidth]{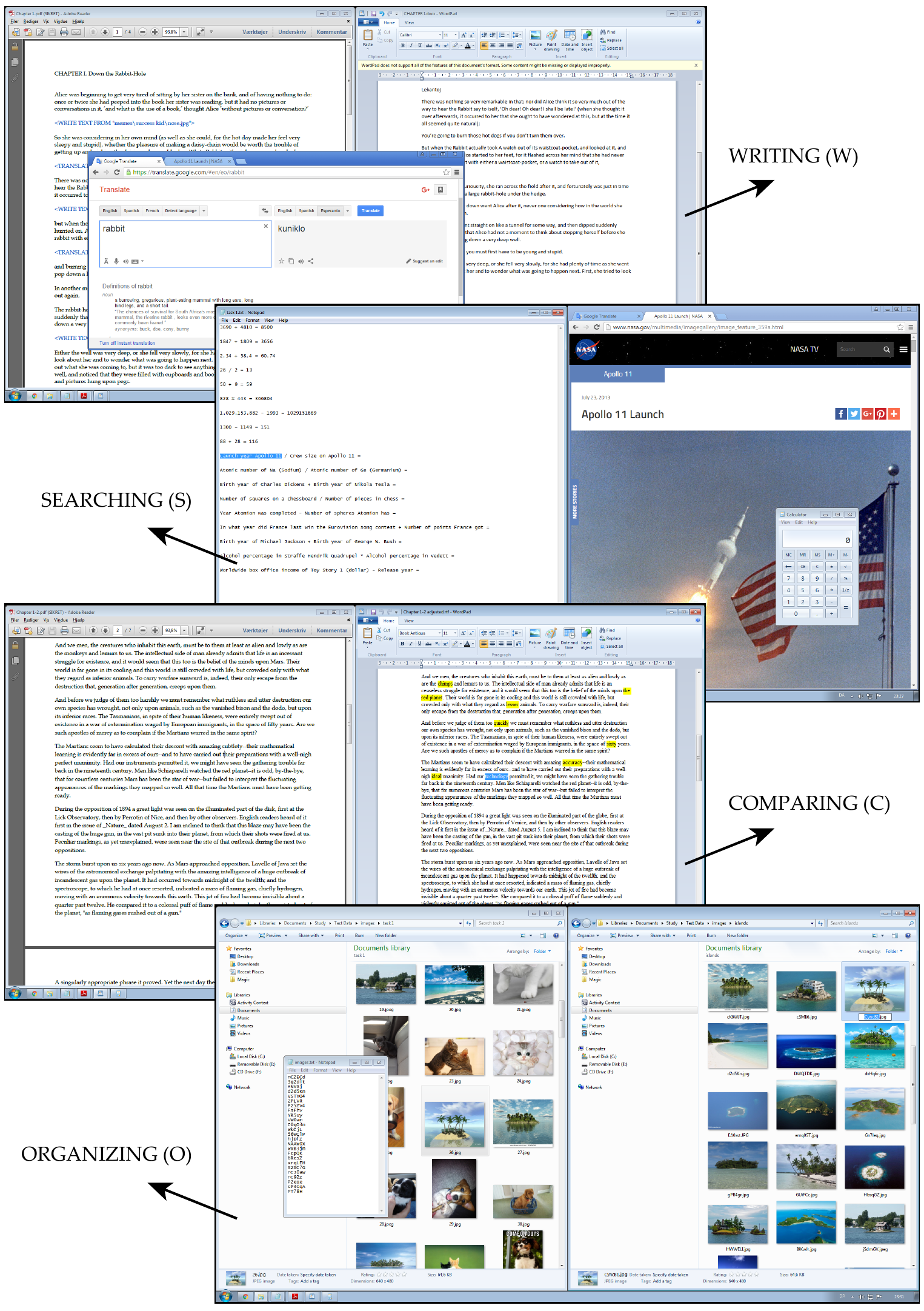}
  \caption{The four tasks part of the experiment.}
  \label{fig:tasks}
\end{figure}

Participants had to switch between four tasks (Fig.~\ref{fig:tasks}), each requiring 2--4 application windows, over the course of 50 minutes.
This is modeled after observations showing that on average knowledge workers work on 12.81 tasks per day~\cite{gonzalez2004constant}, with an average of 2.5 documents open for each (returned-to) task~\cite{czerwinski2004diary}.
This procedure thereby supports our main intent: \emph{simulating heavy multitasking during which several application windows are open simultaneously}.
We used the same tasks as in a recent study on window management~\cite{jeuris2016dedicated} and did not simulate or modify any of the applications used.
The documents required for the tasks were placed in the default ``Documents'' folder.
For each task a text file with a short assignment description was provided, referring to the required files and folder location needed for that task.
No other documents needed to be accessed during the study.

\begin{description}
\item[Writing (W)] Participants type text found within a PDF file into a new text document (copy is disabled). No formatting needs to be applied.
At regular intervals the text includes assignments asking participants to either substitute the assignment with text displayed in an image pointed to on the hard drive, or with the translation of a word using Google Translate. Image thumbnails are available.
This implies \emph{four} applications need to be used: a file explorer, a browser, PDF reader, and a text processor.

\item[Searching (S)] Participants perform calculations based on searching the Internet. For example, ``Height of the Eiffel Tower in meters + year when it was completed =''. Both the intermediate and final results need to be written in a text file.
This implies \emph{three} applications need to be used: a browser, a text processor, and a calculator.

\item[Comparing (C)] Participants highlight differences between an original and modified text. Modifications include synonyms, left out words, or additional words, but the modified text is still grammatically correct.
This implies \emph{two} applications need to be used: two text processors.

\item[Organizing (O)] To mimic folder navigation, a folder hierarchy contains images organized by type of object (e.g., bridges, islands ...). A task folder contains the same images but disorganized. Participants need to identify what is displayed in the images from the task folder, find the folder of the corresponding type, within it find the image, and subsequently copy its filename into a text document.
This implies \emph{two} applications need to be used: a file explorer and a text processor. However, experienced users generally use several file explorer windows. In addition, the image viewer can be used when image thumbnails are unclear.
\end{description}

\subsection{Procedure}
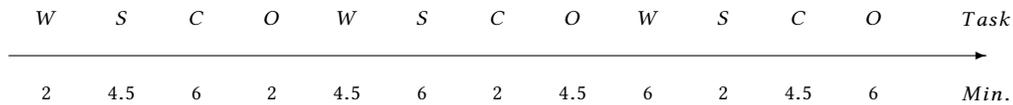
\begin{figure}[!h]
\small
\setlength{\unitlength}{1cm}
\centering
\begin{picture}(14,1)
\put(0,0){\vector(1,0){13}}
\put(0.5,0.5){\makebox(0,0)[c]{$W$}}
\put(1.5,0.5){\makebox(0,0)[c]{$S$}}
\put(2.5,0.5){\makebox(0,0)[c]{$C$}}
\put(3.5,0.5){\makebox(0,0)[c]{$O$}}
\put(4.5,0.5){\makebox(0,0)[c]{$W$}}
\put(5.5,0.5){\makebox(0,0)[c]{$S$}}
\put(6.5,0.5){\makebox(0,0)[c]{$C$}}
\put(7.5,0.5){\makebox(0,0)[c]{$O$}}
\put(8.5,0.5){\makebox(0,0)[c]{$W$}}
\put(9.5,0.5){\makebox(0,0)[c]{$S$}}
\put(10.5,0.5){\makebox(0,0)[c]{$C$}}
\put(11.5,0.5){\makebox(0,0)[c]{$O$}}
\put(13,0.5){\makebox(0,0)[c]{$Task$}}
\put(0.5,-0.5){\makebox(0,0)[c]{$2$}}
\put(1.5,-0.5){\makebox(0,0)[c]{$4.5$}}
\put(2.5,-0.5){\makebox(0,0)[c]{$6$}}
\put(3.5,-0.5){\makebox(0,0)[c]{$2$}}
\put(4.5,-0.5){\makebox(0,0)[c]{$4.5$}}
\put(5.5,-0.5){\makebox(0,0)[c]{$6$}}
\put(6.5,-0.5){\makebox(0,0)[c]{$2$}}
\put(7.5,-0.5){\makebox(0,0)[c]{$4.5$}}
\put(8.5,-0.5){\makebox(0,0)[c]{$6$}}
\put(9.5,-0.5){\makebox(0,0)[c]{$2$}}
\put(10.5,-0.5){\makebox(0,0)[c]{$4.5$}}
\put(11.5,-0.5){\makebox(0,0)[c]{$6$}}
\put(13,-0.5){\makebox(0,0)[c]{$Min.$}}
\end{picture}
\vspace{0.7cm}
\caption{Task sequence of the writing (W), searching (S), comparing (C), and organizing (O) task.}
\label{fig:task-sequence}
\end{figure}

The study comprised three distinct phases: briefing, training, and evaluation.
During the briefing participants filled out a questionnaire assessing their overall computer literacy and the degree of multitasking they engage in on average.
Meanwhile, the notebook was set up to reflect the users' preferences, including their preferred web browser. 
During a 10 minute training session participants familiarized themselves with a training task set, similar to the tasks used for the real evaluation, and were asked to work on them until they understood what each task entailed.
During the main evaluation, the experimenter notified the user 12 times of when to switch between tasks over the course of 50 minutes. These notifications were done at predetermined intervals of 2, 4.5, and 6 minutes, totaling in 12.5 minutes of work per task (Fig.~\ref{fig:task-sequence}). 
The experimenter announced task switches by stating the required task number and a short description, e.g., ``Now please switch to task A, which is the writing task.''
Participants were instructed that they were allowed to wrap up ongoing work by finishing the current subtask, e.g., a copy/paste operation, or finish writing a sentence, but were not allowed to commence work on a new subtask.
Finishing subtasks was allowed to control for disruptive effects due to differences within the disengagement stage, since interruptions at task boundaries are know to be less disruptive~\cite{bailey2006need}.  

\subsection{Data Analysis}
\begin{figure}
  \centering
  \includegraphics[width=0.98\linewidth]{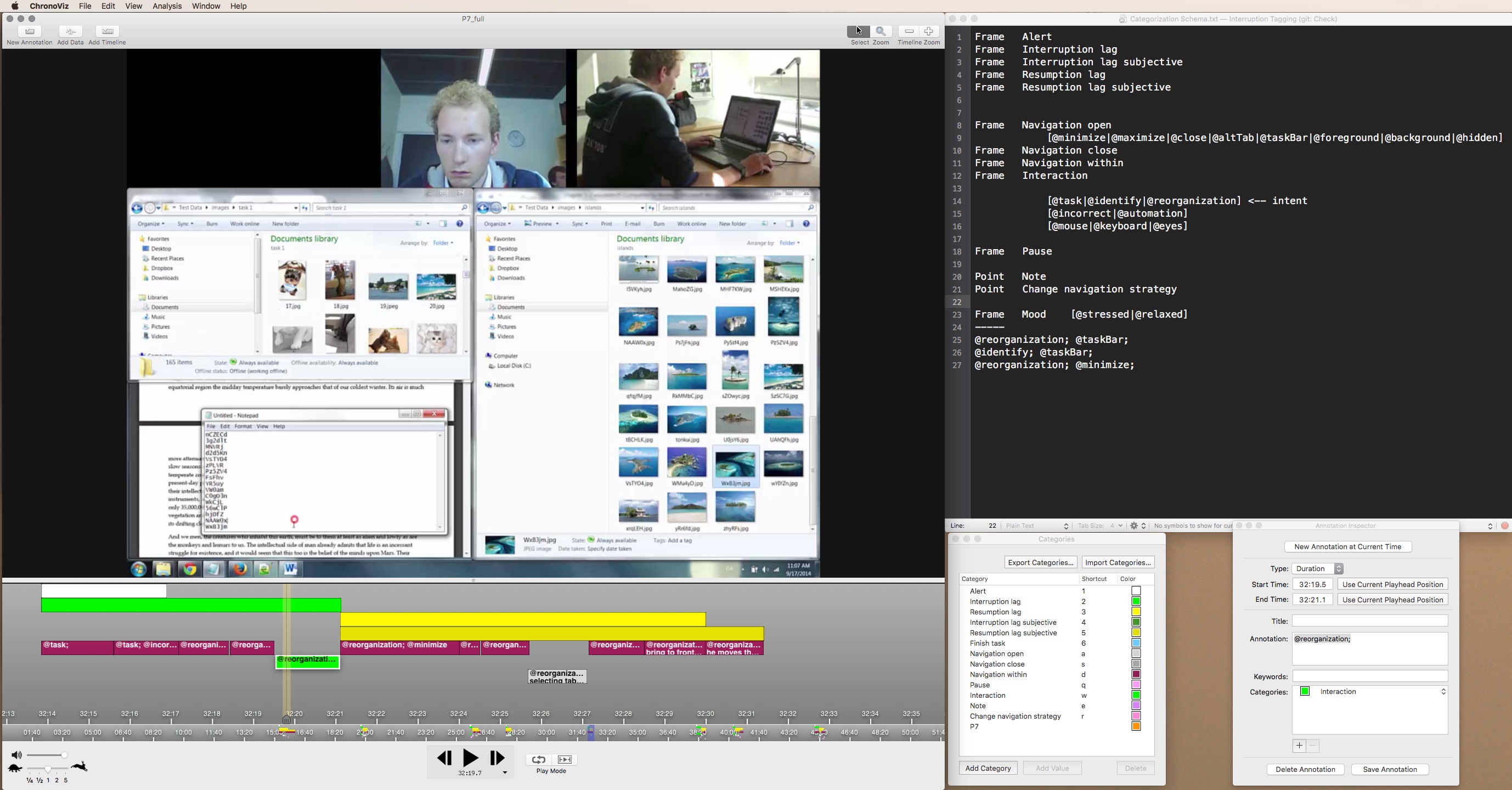}
  \caption{Setup for data analysis in ChronoViz~\protect\cite{fouse2011chronoviz}. The analysis shown is also depicted in Fig.~\protect\ref{fig:detailedbreakdown}, [5] and in the supplementary video.}
  \label{fig:setup}
\end{figure}

Two researchers collaboratively reviewed the video recordings on a large wall display. 
ChronoViz~\cite{fouse2011chronoviz} was used to synchronize, play back, and annotate the data streams. 
This setup allowed simultaneous observation of posture, verbalizations, facial expressions, and interactions of the participants during task switching (Fig.~\ref{fig:setup}). Other segments were ignored.
Throughout several iterations, a thematic coding framework emerged, providing a full breakdown of task switches based on discrete time intervals (Table~\ref{fig:categories}).
All task switches were reinspected until no further changes to the framework needed to be made to account for all observed interactions.
Using the resulting \emph{classification of window management operations}, a third researcher independently validated all annotated data, leading to revisions (after further discussion) where annotations were unclear, or considered inconsistent with the classification.

Annotation consisted of two main phases. 
First, for each task switch three points in time were identified: (i) the beginning of the alert, identified by the voice of the experimenter (Fig.~\ref{fig:firstPage}, \emph{Interruption}); (ii) the start of the task switch, identified by the first interaction aimed at preparing the environment for the next task (e.g., minimizing a window, or moving the cursor towards the taskbar to open a new resource) (Fig.~\ref{fig:firstPage}, \emph{Switch}); and (iii) the start of work on the next task, traditionally associated with the first recordable input users perform on the task set (e.g., a tower of Hanoi move~\cite{monk2008effect})---we call this \emph{objective resumption} (Fig.~\ref{fig:firstPage}, \emph{Resumption}).
However, our observations showed that objective resumption does not always coincide with the time at which users noticeably start working on the next task. Therefore, we included a more subjective measure---\emph{subjective resumption}---also taking observable mental readiness into account.
Example indications for this are: users' verbal utterances (e.g., ``oh yes, here I was!''), physical movements (e.g., fingers used as pointers, relaxing before starting a task, facial expressions), and cursor movements (e.g., hovering over a relevant point in the task set).

Second, the identified task switch intervals were fully annotated according to the classification (Table~\ref{fig:categories}) by playing back the recording at \nicefrac{1}{4} of the original speed. 
During a first pass, task switch intervals were broken down into intervals denoting interactions with \emph{resources} of the system (`Action' categories).
Resources were defined as containers of content which can be shown or hidden separately (notice this definition is interface-agnostic): e.g., files, folders, and browser tabs. During a second pass, an intent and feature of the user interface used was assigned to each of these interactions (`Intent' and `Interface' categories respectively).
Lastly, an additional attribute allowed highlighting whether or not an action could be considered erroneous, given the intent (`Modifier' category).
\section{Results}
\label{sec:results}

\begin{table}
	\centering
	\classificationschema
	\caption{Classification of window management operations, used to analyze task switches when using a window manager.}
	\label{fig:categories}
	\vspace{-1em}
\end{table}

\begin{sidewaysfigure}
\vspace*{15cm}  
	\includegraphics[width=\columnwidth]{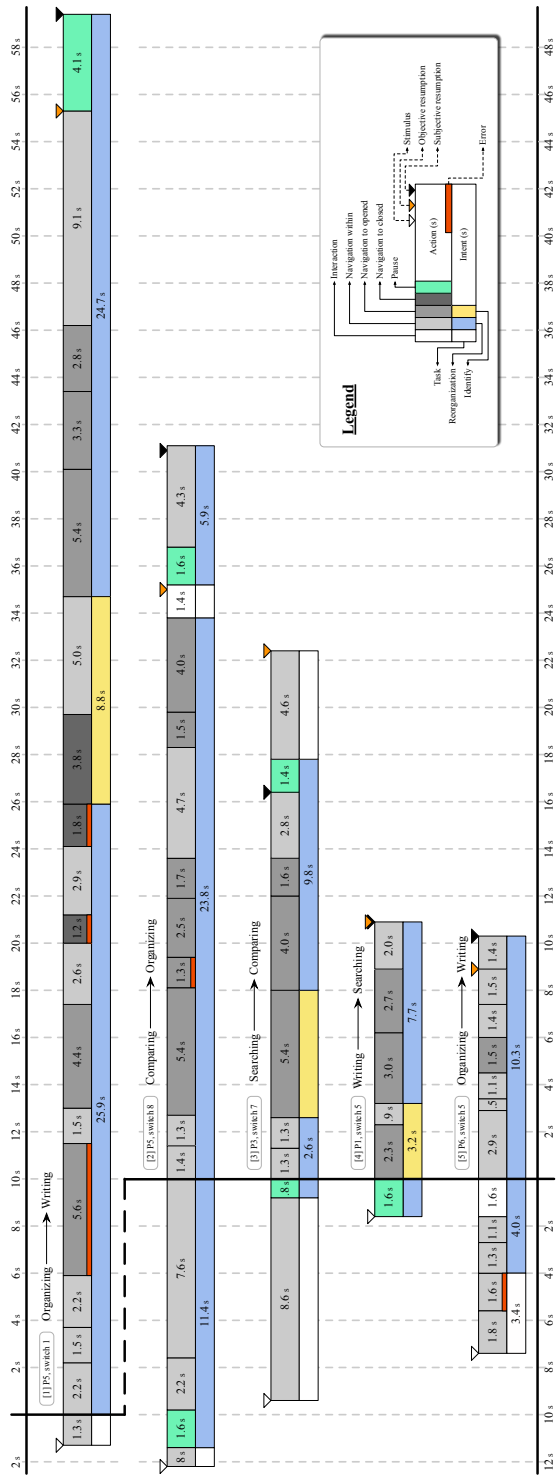}
   \caption{Detailed breakdown of representative task switches for recurring observations. One interesting finding is a frequent pause right before the resumption stage (e.g, in switch [3] and [4]) and a short pause after all navigations to open resources have completed (e.g., switch [2]).}
  \label{fig:detailedbreakdown}
\end{sidewaysfigure}

The outcome of our data analysis is a full breakdown of all observed task switches, following the classification of window management operations we developed throughout several iterations of thematic coding (Table~\ref{fig:categories}).
We visualize the full coding of five representative task switches which we will refer to while presenting recurring observations (Fig.~\ref{fig:detailedbreakdown}).
The results presented here demonstrate how this classification can be used to quantify how well a window manager supports task switching and how such a detailed analysis can lead to interesting new insights.

There are 12 task switches during the evaluation (Fig.~\ref{fig:task-sequence}). During the initial four, participants were asked to start working on a new task; in the remaining eight, they had to return to a previously constructed task.
Therefore, we did not include the initial four task switches in the data analysis as they are not representative of task resumption.
Data of one of the participants (Px) had to be dropped due to a misunderstanding of the experiment which resulted in non-comparable data to the other participants: he did not resume tasks, rather restarted them from the beginning at each task switch. P4 initially did not anticipate having to return to tasks, hence closed task resources. We therefore do not consider his first four task resumptions (task switch 5 through 8).
This results in a total of 44 annotated task switches representative of task resumption, covering 27 minutes and 11 seconds of data (out of five hours of recorded material of participants working on the tasks).

\begin{figure}[b]
  \centering
  \includegraphics[width=0.65\columnwidth]{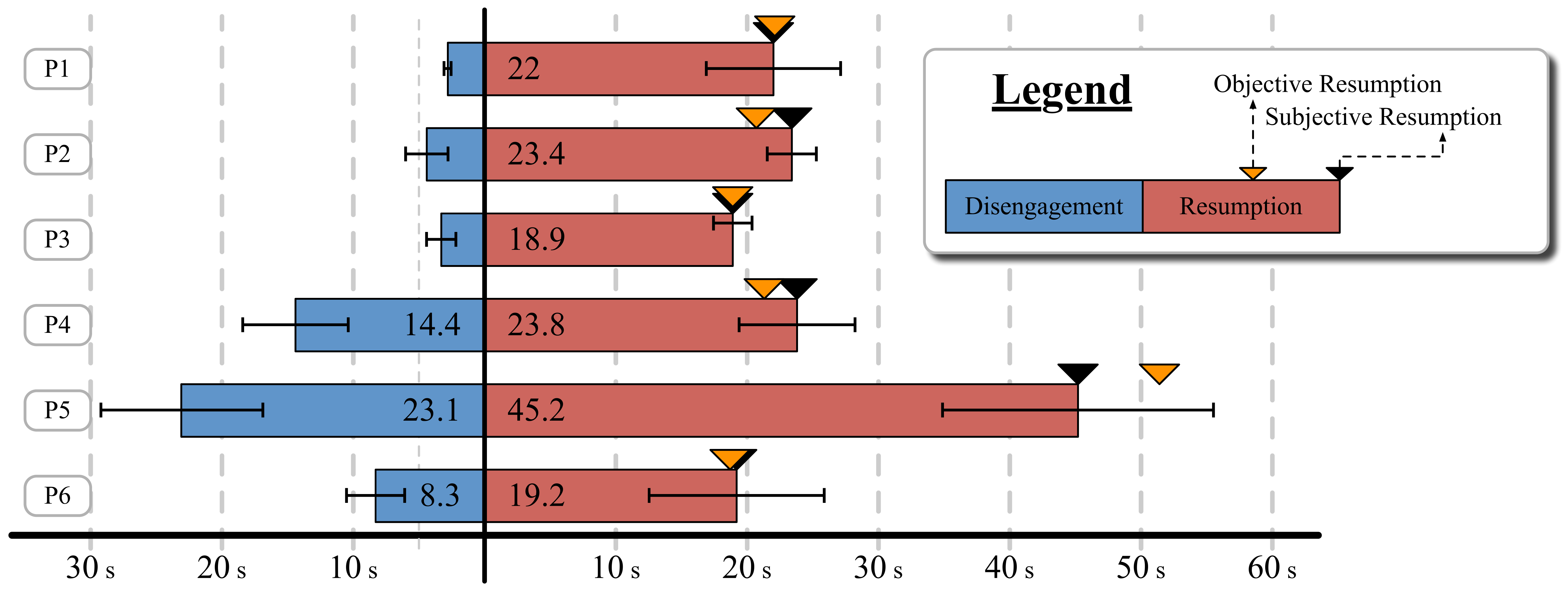}
  \caption{Average disengagement and resumption times per participant.}
  \label{fig:overallaverage}
\end{figure}

Fig.~\ref{fig:overallaverage} shows the average disengagement and subjective resumption time per participant. On average participants spent 8.9 s when suspending tasks ($SD=9.8$, $Min=0.7$, $Max=43.5$) and 25.5 s when resuming tasks ($SD=15.0$, $Min=8.0$, $Max=72.3$).
In addition to subjective resumption time, the difference with average objective resumption time is shown. On average, objective and subjective resumption time differ by 4.8 s ($SD=6.6$, $Min=0.0$, $Max=36.7$). Objective resumption time can exceed subjective resumption time when participants resume a task (e.g., reading) without interacting directly with any of the documents (e.g., see \emph{Minimizing} strategy in the results section on \nameref{sec:strategies}).

\begin{table}[b]
	\centering
	\errors
	\caption{Time and percentage of total task switch time spent on errors for each participant.}
	\label{fig:errors}
	\vspace{-2em}
\end{table}

\begin{sidewaysfigure}
\vspace*{15cm}
  \includegraphics[width=\textwidth]{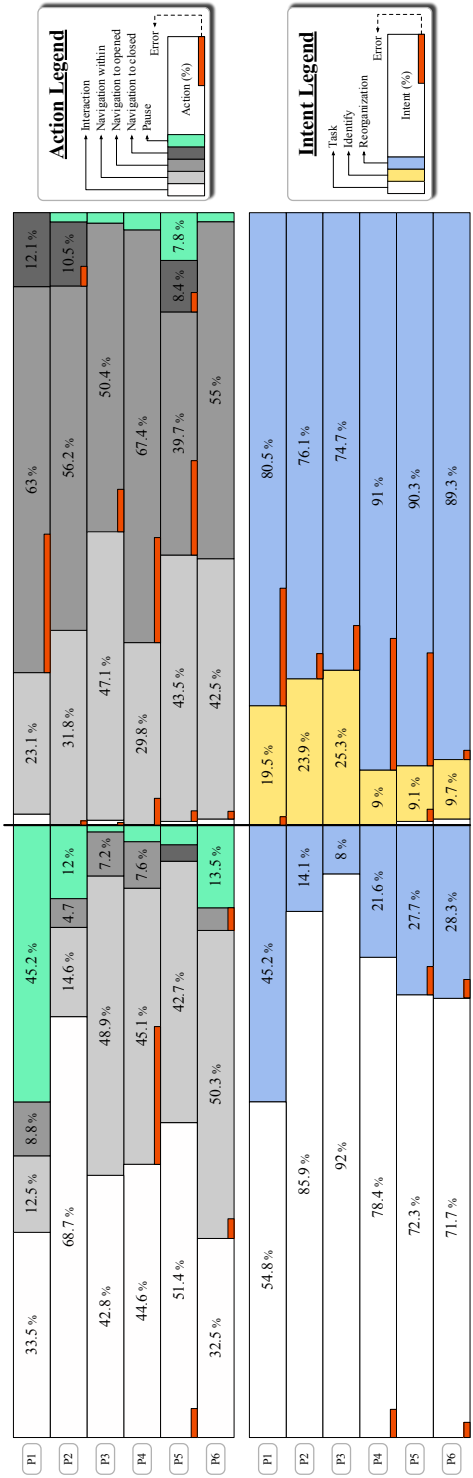}
  \caption{Percentile breakdown of actions and intents of all eight task switches per participant (except just four for P4), for both the disengagement and resumption stage. Half of the task switches contained erroneous actions, mainly impacting navigation to open resources during the resumption stage.}
  \label{fig:percbreakdown}
\end{sidewaysfigure}

Fig.~\ref{fig:percbreakdown} shows a percentile breakdown of actions and intents of all task switches per participant for both the disengagement and resumption stage.
We dropped time intervals where objective resumption time exceeded subjective resumption time, as we deem them not part of the task switch.
There are no actions with the intent of identifying the task during the disengagement stage, as this would indicate the start of the resumption stage. Similarly, during the resumption stage there are almost no actions with the intent of working on the task (only 2 short instances), since such a first action generally denotes the end of the task switch and full resumption of the new task.

In addition, Fig.~\ref{fig:percbreakdown} highlights the percentage of erroneous time intervals per type of action and intent (e.g., retrieving an incorrect resource, saving an incorrect file). Half of the task switches (22 out of 44) contained one or more erroneous actions.
Table~\ref{fig:errors} shows the distribution of error instances across participants.
Eight out of 43 errors could be attributed to users performing ingrained actions part of task switching, with limited observable attention. Most notably, P1 failed six times in opening a resource when using Alt-Tab, after which he fell back to using the taskbar. In one occasion while minimizing windows, P3 used a shortkey operation to save a text file which did not need any saving as it not only was unmodified, but also did not need to be modified as part of the experiment. Lastly, P4 intentively investigated window thumbnails when using the taskbar, only to realize later he did not know which resource he was looking for.
The remaining errors mainly involved resuming incorrect resources (noticeable in navigation to opened) and typos or erroneous clicks (noticeable in interaction).

\subsection{Actions and Intents}
\begin{figure}[h]
  \centering
	\includegraphics[width=0.65\columnwidth]{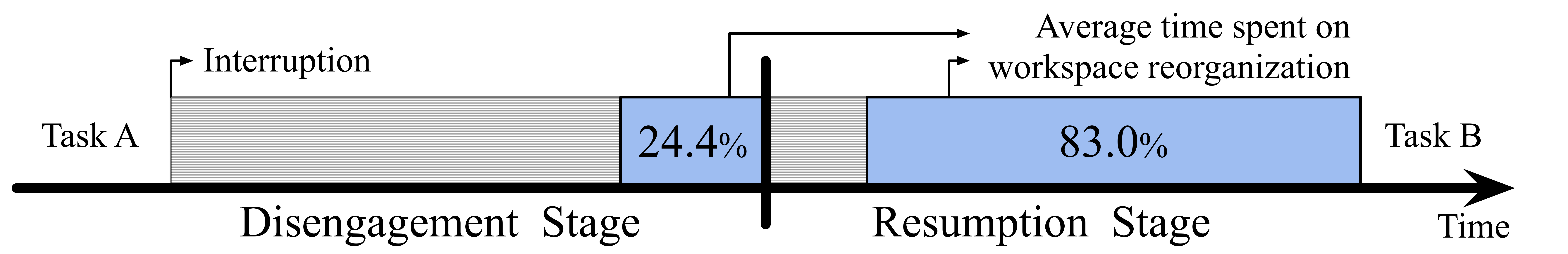}
  \caption{The average percentage of time spent during task switching on \emph{reorganizing} the workspace by our participants.}
  \label{fig:average_reorganization}
\end{figure}

With a high-level overview of the makeup of disengagement and resumption time in place, it is now possible to describe what these containing actions and intents entail. The main interest of this study is to gain a better understanding of how window managers support \emph{reorganizing} the workspace (Fig.~\ref{fig:average_reorganization}).
Therefore, these are the main time intervals which will be presented in the subsequent results.
Table~\ref{fig:breakdownTable} lists the overall percentages as part of \emph{reorganization work} for each of the actions in our classification for both the disengagement ($\Box$) and resumption ($\triangleright$) stage, thus excluding actions with the intent to identify or to work on tasks.

\begin{table}[h]
	\centering
	\breakdowns
	\caption{The makeup of reorganization work during disengagement ($\Box$) and resumption ($\triangleright$). Rare instances: $^i$}
	\label{fig:breakdownTable}
\end{table}

\subsubsection{Disengagement stage}
After having received the alert to switch to the next task, participants generally finished up their current subtask (e.g., Fig.~\ref{fig:detailedbreakdown}, [3]).
In just one instance this did not occur (Fig.~\ref{fig:detailedbreakdown}, [4]), as P1 had just finished a subtask when the alert was given.
Participants spent the remaining disengagement time on reorganizing the task set for future resumption (e.g., saving resources, renaming files, highlighting text, and leaving notes in text files) or taking a break (e.g., Fig.~\ref{fig:detailedbreakdown}, [2]).
In 15 cases there was no reorganization work, thus disengagement constituted solely wrapping up work on the previous task (e.g., Fig.~\ref{fig:detailedbreakdown}, [1]).

Reorganization work within the disengagement stage is largely composed of actions facilitating future resumption of the task (51.6\%), namely interactions with and navigation within resources.
Interactions mainly reflect save and rename operations.
Navigation within reflects users repositioning, resizing, and scrolling within windows in order to position them for later resumption.
However, only half of the participants (P4, P5, P6) prepared tasks for future resumption on a regular basis; P1 never did, and P2 and P3 only on a single occasion.
All participants (P1 in particular) regularly took breaks during the disengagement stage (46.6\%), which lasted on average \mbox{1.8 s} per task switch ($N=18$, $SD=1.2$, $Min=0.7$, $Max=5.9$).
In particular, 14 out of the 44 task switches ended with a short break prior to the start of the task switch (e.g., Fig.~\ref{fig:detailedbreakdown}, [3]). This break was generally associated with a change in posture (leaning backwards) and temporarily closing one's eyes.
Navigation to opened occurred in just one instance (\mbox{2.2 s}) during reorganization work in the disengagement stage. P6 considered opening a resource to resume the next task, but aborted this action in order to write one final character acting as a placeholder for future resumption of the task.

\subsubsection{Resumption stage}
At the start of the resumption stage many participants could not recall the exact resources they needed in order to resume the task (31 out of 44 task switches), in which case they commonly would reopen the task description.
When this occurred, it took on average 6.0 s before participants actually continued resuming the required task ($N=31$, $SD=3.4$, $Min=1.9$, $Max=18.8$).
The remaining time (disregarding a few short task intervals) was spent on setting up the workspace to start work on the new task (reorganization averages shown in Table~\ref{fig:breakdownTable}).
This took in total on average 21.3 s ($N=44$, $SD=13.2$, $Min=7.8$, $Max=54.6$).

On average, navigation to opened occurred 2.9 times per task switch ($SD=0.9$) with the purpose of reorganizing the workspace during task resumption, taking up 10.3 s in total ($N=44$, $SD=5.2$, $Min=1.5$, $Max=26.4$); navigating within resources 3.6 times ($SD=1.9$), taking up 8.8 s in total ($N=43$, $SD=8.7$, $Min=1.3$, $Max=41.1$).
When pauses occurred (14 out of 44 task switches), they were always indicative of having restored the necessary resources, and were followed by navigating within the retrieved resource to find the location where to continue work (e.g., Fig.~\ref{fig:detailedbreakdown}, [2]). These pauses on average lasted 2.6 s ($N=14$, $SD=2.2$, $Min=0.2$, $Max=7.7$).
In a few instances, participants closed resources as part of ongoing work and had to reopen them during the resumption stage using either Windows explorer or by reopening browser windows or tabs. Navigation to closed took up 7.2~s in total on average ($N=10$, $SD=3.6$, $Min=3.0$, $Max=13.5$).
Lastly, just three actions during the resumption stage represented interactions. During these, files were saved (0.5~s, 0.6~s) and renamed (6.2~s).

\subsection{Window Management Features}
\label{sec:strategies}
All actions marked as either using Alt-Tab or the taskbar, except one, were marked as navigations to opened resources; one action using the taskbar was categorized as navigation within, as P5 spent a total of five seconds investigating a set of window thumbnails in order to select the right one. We will therefore solely report on which window management features were used as part of navigating to opened resources.
On average, navigating to an opened resource took 3.5~s ($N=165$, $SD=2.4$, $Min=0.5$, $Max=20.0$).
Predominantly, the taskbar was used to this end by all participants (Table~\ref{fig:interface}). Only two users (P1 and P3) occasionally used Alt-Tab. Remarkably, P3 never retrieved a window by selecting a partially visible window from the foreground.
Using the taskbar to open resources on average took 3.8~s ($N=132$, $SD=2.5$, $Min=0.5$, $Max=20.0$); using Alt-Tab 3.2~s ($N=8$, $SD=2.2$, $Min=1.5$, $Max=7.5$); opening from the foreground 2.1~s ($N=25$, $SD=1.1$, $Min=0.9$, $Max=6.2$).

\begin{table}[b]
	\centering
	\winfeatures
	\caption{Features used for navigation to opened.}
	\label{fig:interface}
\end{table}

We observed that the appropriation of window management features during task switching varied greatly from user to user. Participants used a combination of five different strategies.
The first strategy (\textit{Window Movement}: P5) involved extensive repositioning and resizing of windows (e.g., Fig.~\ref{fig:detailedbreakdown}, [2]). This indicated that P5 was seemingly unaware of the window docking feature (docking windows to the side so that they occupy half of the workspace).
In contrast, in a second strategy participants relied on strict side-by-side window configurations using this feature (\textit{Side-by-side}: all except P5).
This simplified the process of switching between tasks by mainly relying on selecting two required windows from the taskbar.
A third strategy involved overlapping windows in various ways, keeping the content of multiple windows partially visible (\textit{Overlapping}: P1, P4, P6).
Irrelevant windows were often minimized, revealing windows for the required task underneath. Later, the taskbar was used to bring them back to the foreground when resuming tasks.
A fourth strategy consisted of minimizing all windows during disengagement (\textit{Minimizing}: P3, P5). Interestingly, participants never used the shortcut key (show desktop) to this end. Possibly users were hoping to find relevant windows underneath (as observed when overlapping windows).
A final observed strategy was to tile windows, placing them at semi-fixed positions (\textit{Tiling}: P6).
While working on a task, there was little overlap among foreground windows, and windows were rarely repositioned. This allowed P6 to keep more than two windows permanently visible (as in the side-by-side strategy) while working on a task.
\section{Discussion}
\label{sec:discussion}

The study reported on above is relevant in two ways. 
First, it provides an assessment of the impact of window management as part of sequential multitasking on a desktop computer---which we will denote as the \emph{hidden cost of window management}.
Second, the classification of window management operations shown in Table~\ref{fig:categories} provides a framework for designers of window management technologies to become sensitized to the different micro operations which need to be supported when switching between tasks.
To demonstrate, we will use this classification to discuss existing \emph{window management techniques} and identify opportunities for design that could further reduce the overhead of switching between tasks.

\subsection{The hidden cost of window management}
In an observational study of knowledge workers, Gonz\'alez and Mark~\cite{gonzalez2004constant} report that the time spent within a specific working sphere (``units of work or activities that people divide their work into on a daily basis'') before switching to another, ranges from 07:41 (mm:ss) to 16:24. 
Considering an 8 hour work day and using our experimental data, we can extrapolate the daily cost of task switching per participant as shown in Table~\ref{fig:cost}.
For this analysis, we disregarded actions with the \emph{intent of working on the task} as they simply represent a continuation of work. Similarly, we disregarded actions in which participants had to \emph{identify} the task resources to use, as they are arguably caused by our participants not being the original owners of the tasks.
As shown in Table~\ref{fig:cost}, P5 is an outlier that would spend 50 minutes per day on task switching when faced with a similar scenario as presented in our study. 
On average, treating P5 as an outlier, the participants in our study would spend between 09:45 and 20:50 of their day reorganizing their workspace as part of multitasking.

\begin{table}[b]
	\centering
	\costPerP
	\caption{Predicted time spent on reorganization based on 16:24 estimate$^{[1]}$ and 7:41 estimate$^{[2]}$.}
	\label{fig:cost}
\end{table}

Novel window management systems consistently refer to task switching as a problematic practice that warrants attention.
These systems typically (and for the most part, rightfully) assume that novel approaches would outperform traditional window managers.
For example, \citeN{bardram2006activitybar} stated: \emph{``We have refrained from measuring task completion time. It is trivial to see that users of [our system] would out-perform standard Windows XP users if asked to perform tasks that require handling parallel tasks, accessing lots of digitial material, [and] coping with frequent interruptions.''}
Only one earlier study has measured task resumption time using a traditional window manager~\cite{jeuris2016dedicated} ($\mu=15.8$ s, $SD=7.0$). Using the same task sets, we replicated this measure (\mbox{$\mu=25.5$ s}, $SD=15.0$) and in addition provided a detailed breakdown of constituting window management operations.
Our classification and results clarify where the variance among different users and task switches may stem from and gives a first impression of the degree to which different window management operations contribute to overall task switch time.

\begin{figure}
  \centering
  \includegraphics[width=\textwidth]{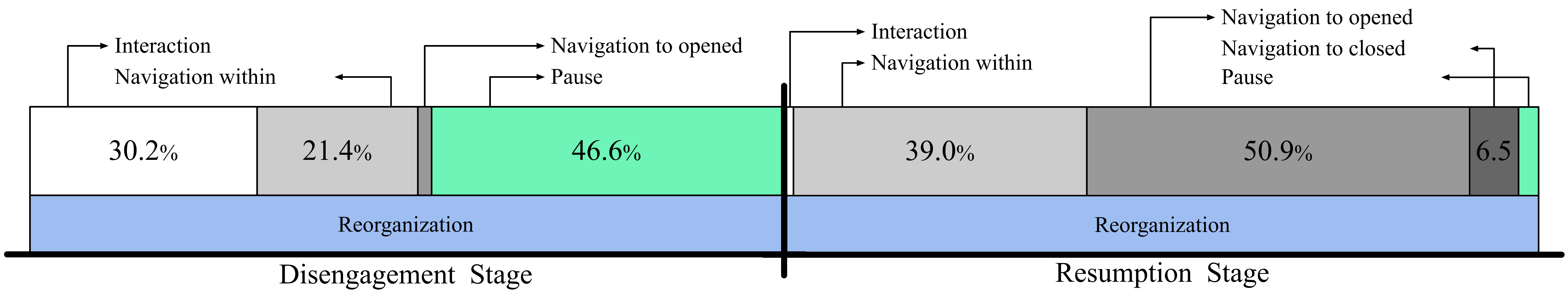}
  \caption{Average breakdown of reorganization actions during task switching by our participants.}
  \label{fig:reorganization_percentages}
\end{figure}

Fig.~\ref{fig:reorganization_percentages} provides an overview of how much each individual action contributed to reorganization work in both the disengagement and resumption stage.
In particular, \emph{navigating to open} and \emph{closed} resources during the resumption stage (57.4\%) does not seem to contribute to the overall goal of the task at hand, is error-prone (22 out of 44 task switches contained one or more errors), and requires a significant amount of concentration.
An interesting pattern we observed emphasizes this finding: navigation to opened and closed resources in many cases was followed by a pause.
Only after taking a break, participants would navigate within documents to find the exact location where to pick up work (e.g., Fig.~\ref{fig:detailedbreakdown}, [2]).
Intrigued by this finding, we reinvestigated the data and could observe very short subtle hints---hence not coded---indicating pauses at the end of most navigate to opened/closed sequences preceding navigations within, e.g., closed eyes and small posture readjustments.
We therefore speculate that the entire sequence of reconstructing the work environment could be considered (and is experienced by users) as a task in and of itself.
Attempting to observe this sequence of events might prove useful for interruption management systems, since interruptions at task boundaries are known to be less disruptive~\cite{bailey2006need}.

We noticed that the purpose behind \emph{navigation within} resources during the disengagement stage (21.4\%) was to leave behind visual markers to facilitate later resumption.
For instance, a common strategy included scrolling within a window so that the next subtask would be displayed first.
For the same reason, participants \emph{interacted} with resources (30.2\%): e.g., selecting text in a PDF file, or typing reminders in text documents part of the task set.
Prior studies have shown that leaving behind such cues reduces the amount of time required to resume a task~\cite{altmann2004task}.
Our observations show examples of the various application-specific techniques that participants rely on in order to leave behind these cues.

In the presence of a significant number of resources, our participants struggled to effectively use the features provided by the operating system. Even though the Windows taskbar aggregates resources by application as a way to reduce taskbar clutter, users on average spent 3.5 s per resource that needed to be opened ($N=165$).
In addition, we noticed participants had a hard time disambiguating between multiple resources belonging to the same application using the taskbar preview.
Regardless, the taskbar seemed the preferred mechanism to switch between tasks. During the experiment, it proved more effective than Alt-Tab, which lists \emph{all} open resources: six times P1 attempted to use Alt-Tab during task resumption, only to cancel the operation when faced with too many resources. However, Alt-Tab is optimized to switch between recently accessed resources, making it likely more useful to switch between windows (belonging to the same task set) while working on a task.

\subsection{Window management techniques}
More recent window managers (e.g., on Mac OS and Windows 10) as well as a number of research-based technologies incorporate more advanced features for window management than the ones tested in our study (e.g., \cite{robertson2004scalable,tashman2012windowscape,henderson1986rooms,agarawala2006bumptop}).
The support for multitasking in these window management systems can be analyzed and categorized in terms of how they \emph{reduce the time for}, or \emph{reduce the number of} specific operations identified in our window management classification (Table~\ref{fig:categories}).

Of all window management operations, \emph{navigation to closed} (in case no application window for the resource is open) is the most costly operation.
Likely, this is why it was targeted by early innovations.
File shortcuts, bookmarks, and advanced search can reduce the (\emph{reorganization}) time needed to navigate to closed resources.
To help \emph{identify} the necessary resource, previews can show resources prior to opening them (e.g., thumbnails or larger previews of selected items), further reducing the time needed to navigate to closed resources.
But, by increasing the available display space (most knowledge workers now have a high resolution screen or dual-monitor setup~\cite{ling2017twobetter}) more resources can remain open without cluttering the workspace, thereby replacing \emph{navigation to closed} operations with \emph{navigation to opened}---operations which are much more efficient.

Once it is possible to keep more resources open, a new bottleneck arises: having to identify a resource among a large set of open windows slows down \emph{navigation to opened}.
Typically, an overview is made available which represents all open resources, but as demonstrated in our study (the use of Alt-Tab), this does not scale well.
Although more modern interpretations (such as macOS Expos\' e and Windows 10 task view) improve on this, the underlying problem remains the same.
Commonly, the next step is to support \emph{resource grouping}:
the system either automates or provides mechanisms to group related resources together and retrieve them as a whole (e.g., virtual desktops~\cite{henderson1986rooms}, browser tabs, or grouping windows in the margin~\cite{robertson2004scalable}), thereby reducing the \emph{number} of required navigations to opened resources and providing users with additional cues on where the resource they are looking for might be located. 
Within research, some `activity-centric computing' systems take this one step further~\cite{bardram2006activitybar,voida2008reframing,houben2013cam,jeuris2014laevo}. Blurring the lines between \emph{navigation to opened} and \emph{navigation to closed}, activity-centric computing systems aim to support computational, stateful, `activities'---including the associated window configurations---which can be suspended and resumed at any given time (e.g., represented as `tasks' on a time line~\cite{jeuris2014laevo}). The goal is to be able to close any group of related resources and reopen the same workspace weeks or even months later after the activity has been suspended.
Most recently, this functionality has been announced and made available on insider builds of Windows 10, named `sets'. Sets are stateful collections of resources accessible from a timeline, represented as tabs in a single application window (in contrast to activity-centric computing systems which typically represents activities as virtual desktops). 

\emph{Navigation within} resources (e.g., scrolling, moving, or resizing) seems to be the least explored operation as part of switching between tasks, and is therefore a suitable topic for further research.
The visual cues left behind by our participants indicate an interesting opportunity for design: rather than relying on application-specific mechanisms to leave behind such cues, cross-application functionality could be introduced (i.e., handled by the operating system/window manager) to facilitate the resumption of suspended tasks. For example, the ability to add annotations on any type of resource, either through freehand drawings or by attaching post-it notes.
Further inspiration can be drawn from common visualization and navigation techniques: `focus-plus-context' reduces the time spent on going back and forth within a document by providing access to both detailed content and an overview to navigate to other locations, and `semantic navigation' allows to quickly jump to meaningful positions within a document based on the current position in the document (e.g., jumping to the end of a code block in a development environment).

\section{Conclusion}
This paper presents detailed findings on window management `micro operations' as part of sequential multitasking.
So far, this topic has received little to no attention in research, even though window managers have been around for over four decades.
To support this type of research, we introduced a \emph{classification of window management operations},
which can be used to analyze, quantify, and understand how the operations involved in multitasking are supported by different types of window managers.
This provides the basic building blocks to systematically compare different window management techniques and how they impact the user.
We demonstrate the utility of this classification by (i) analyzing the `multitasking profile' of six participants with varying levels of expertise with Windows 7, both quantitatively and qualitatively, and by (ii) using it to structure a short review of window management techniques, highlighting future areas for research.

\bibliographystyle{ACM-Reference-Format}
\bibliography{references}


\begin{thebibliography}{48}


\ifx \showCODEN    \undefined \def \showCODEN     #1{\unskip}     \fi
\ifx \showDOI      \undefined \def \showDOI       #1{#1}\fi
\ifx \showISBNx    \undefined \def \showISBNx     #1{\unskip}     \fi
\ifx \showISBNxiii \undefined \def \showISBNxiii  #1{\unskip}     \fi
\ifx \showISSN     \undefined \def \showISSN      #1{\unskip}     \fi
\ifx \showLCCN     \undefined \def \showLCCN      #1{\unskip}     \fi
\ifx \shownote     \undefined \def \shownote      #1{#1}          \fi
\ifx \showarticletitle \undefined \def \showarticletitle #1{#1}   \fi
\ifx \showURL      \undefined \def \showURL       {\relax}        \fi
\providecommand\bibfield[2]{#2}
\providecommand\bibinfo[2]{#2}
\providecommand\natexlab[1]{#1}
\providecommand\showeprint[2][]{arXiv:#2}

\bibitem[\protect\citeauthoryear{Adler and Benbunan-Fich}{Adler and
  Benbunan-Fich}{2012}]%
        {adler2012juggling}
\bibfield{author}{\bibinfo{person}{Rachel~F Adler} {and}
  \bibinfo{person}{Raquel Benbunan-Fich}.} \bibinfo{year}{2012}\natexlab{}.
\newblock \showarticletitle{Juggling on a high wire: Multitasking effects on
  performance}.
\newblock \bibinfo{journal}{\emph{International Journal of Human-Computer
  Studies}} \bibinfo{volume}{70}, \bibinfo{number}{2} (\bibinfo{year}{2012}),
  \bibinfo{pages}{156--168}.
\newblock


\bibitem[\protect\citeauthoryear{Adler and Benbunan-Fich}{Adler and
  Benbunan-Fich}{2013}]%
        {adler2013self}
\bibfield{author}{\bibinfo{person}{Rachel~F. Adler} {and}
  \bibinfo{person}{Raquel Benbunan-Fich}.} \bibinfo{year}{2013}\natexlab{}.
\newblock \showarticletitle{Self-interruptions in discretionary multitasking}.
\newblock \bibinfo{journal}{\emph{Computers in Human Behavior}}
  \bibinfo{volume}{29}, \bibinfo{number}{4} (\bibinfo{year}{2013}),
  \bibinfo{pages}{1441--1449}.
\newblock
\showISSN{0747-5632}
\urldef\tempurl%
\url{https://doi.org/10.1016/j.chb.2013.01.040}
\showDOI{\tempurl}


\bibitem[\protect\citeauthoryear{Adler and Benbunan-Fich}{Adler and
  Benbunan-Fich}{2015}]%
        {adler2015effects}
\bibfield{author}{\bibinfo{person}{Rachel~F Adler} {and}
  \bibinfo{person}{Raquel Benbunan-Fich}.} \bibinfo{year}{2015}\natexlab{}.
\newblock \showarticletitle{The effects of task difficulty and multitasking on
  performance}.
\newblock \bibinfo{journal}{\emph{Interacting with Computers}}
  \bibinfo{volume}{27}, \bibinfo{number}{4} (\bibinfo{year}{2015}),
  \bibinfo{pages}{430--439}.
\newblock


\bibitem[\protect\citeauthoryear{Agarawala and Balakrishnan}{Agarawala and
  Balakrishnan}{2006}]%
        {agarawala2006bumptop}
\bibfield{author}{\bibinfo{person}{Anand Agarawala} {and}
  \bibinfo{person}{Ravin Balakrishnan}.} \bibinfo{year}{2006}\natexlab{}.
\newblock \showarticletitle{Keepin' It Real: Pushing the Desktop Metaphor with
  Physics, Piles and the Pen}. In \bibinfo{booktitle}{\emph{Proc. CHI}}.
  \bibinfo{publisher}{ACM}, \bibinfo{pages}{1283--1292}.
\newblock
\showISBNx{1-59593-372-7}
\urldef\tempurl%
\url{https://doi.org/10.1145/1124772.1124965}
\showDOI{\tempurl}


\bibitem[\protect\citeauthoryear{Altmann and Trafton}{Altmann and
  Trafton}{2004}]%
        {altmann2004task}
\bibfield{author}{\bibinfo{person}{E.~M. Altmann} {and} \bibinfo{person}{J.~G.
  Trafton}.} \bibinfo{year}{2004}\natexlab{}.
\newblock \showarticletitle{Task interruption: Resumption lag and the role of
  cues}. In \bibinfo{booktitle}{\emph{Proc. CogSci}}.
  \bibinfo{publisher}{Lawrence Erlbaum Associates}, \bibinfo{pages}{43--48}.
\newblock


\bibitem[\protect\citeauthoryear{Bailey and Konstan}{Bailey and
  Konstan}{2006}]%
        {bailey2006need}
\bibfield{author}{\bibinfo{person}{Brian~P Bailey} {and}
  \bibinfo{person}{Joseph~A Konstan}.} \bibinfo{year}{2006}\natexlab{}.
\newblock \showarticletitle{On the need for attention-aware systems: Measuring
  effects of interruption on task performance, error rate, and affective
  state}.
\newblock \bibinfo{journal}{\emph{Computers in Human Behavior}}
  \bibinfo{volume}{22}, \bibinfo{number}{4} (\bibinfo{year}{2006}),
  \bibinfo{pages}{685--708}.
\newblock


\bibitem[\protect\citeauthoryear{Bailey, Konstan, and Carlis}{Bailey
  et~al\mbox{.}}{2001}]%
        {bailey2001effects}
\bibfield{author}{\bibinfo{person}{Brian~P Bailey}, \bibinfo{person}{Joseph~A
  Konstan}, {and} \bibinfo{person}{John~V Carlis}.}
  \bibinfo{year}{2001}\natexlab{}.
\newblock \showarticletitle{The effects of interruptions on task performance,
  annoyance, and anxiety in the user interface}. In
  \bibinfo{booktitle}{\emph{Proc. INTERACT}}. \bibinfo{publisher}{Springer},
  \bibinfo{pages}{593--601}.
\newblock


\bibitem[\protect\citeauthoryear{Bannon, Cypher, Greenspan, and Monty}{Bannon
  et~al\mbox{.}}{1983}]%
        {bannon1983activity}
\bibfield{author}{\bibinfo{person}{Liam Bannon}, \bibinfo{person}{Allen
  Cypher}, \bibinfo{person}{Steven Greenspan}, {and}
  \bibinfo{person}{Melissa~L. Monty}.} \bibinfo{year}{1983}\natexlab{}.
\newblock \showarticletitle{Evaluation and Analysis of Users' Activity
  Organization}. In \bibinfo{booktitle}{\emph{Proc. CHI}}.
  \bibinfo{publisher}{ACM}, \bibinfo{pages}{54--57}.
\newblock
\showISBNx{0-89791-121-0}
\urldef\tempurl%
\url{https://doi.org/10.1145/800045.801580}
\showDOI{\tempurl}


\bibitem[\protect\citeauthoryear{Bardram, Bunde-Pedersen, and Soegaard}{Bardram
  et~al\mbox{.}}{2006}]%
        {bardram2006activitybar}
\bibfield{author}{\bibinfo{person}{Jakob Bardram}, \bibinfo{person}{Jonathan
  Bunde-Pedersen}, {and} \bibinfo{person}{Mads Soegaard}.}
  \bibinfo{year}{2006}\natexlab{}.
\newblock \showarticletitle{Support for Activity-based Computing in a Personal
  Computing Operating System}. In \bibinfo{booktitle}{\emph{Proc. CHI}}.
  \bibinfo{publisher}{ACM}, \bibinfo{pages}{211--220}.
\newblock
\showISBNx{1-59593-372-7}
\urldef\tempurl%
\url{https://doi.org/10.1145/1124772.1124805}
\showDOI{\tempurl}


\bibitem[\protect\citeauthoryear{Beale and Edmondson}{Beale and
  Edmondson}{2007}]%
        {beale2007multiple}
\bibfield{author}{\bibinfo{person}{Russell Beale} {and}
  \bibinfo{person}{William Edmondson}.} \bibinfo{year}{2007}\natexlab{}.
\newblock \showarticletitle{Multiple Carets, Multiple Screens and
  Multi-Tasking: New Behaviours with Multiple Computers}. In
  \bibinfo{booktitle}{\emph{Proc. BCS HCI}}. British Computer Society,
  \bibinfo{pages}{55--64}.
\newblock


\bibitem[\protect\citeauthoryear{Boehm-Davis and Remington}{Boehm-Davis and
  Remington}{2009}]%
        {boehm2009reducing}
\bibfield{author}{\bibinfo{person}{Deborah~A Boehm-Davis} {and}
  \bibinfo{person}{Roger Remington}.} \bibinfo{year}{2009}\natexlab{}.
\newblock \showarticletitle{Reducing the disruptive effects of interruption: A
  cognitive framework for analysing the costs and benefits of intervention
  strategies}.
\newblock \bibinfo{journal}{\emph{Accident Analysis \& Prevention}}
  \bibinfo{volume}{41}, \bibinfo{number}{5} (\bibinfo{year}{2009}),
  \bibinfo{pages}{1124--1129}.
\newblock


\bibitem[\protect\citeauthoryear{Brumby, Cox, Back, and Gould}{Brumby
  et~al\mbox{.}}{2013}]%
        {brumby2013recovering}
\bibfield{author}{\bibinfo{person}{Duncan~P Brumby}, \bibinfo{person}{Anna~L
  Cox}, \bibinfo{person}{Jonathan Back}, {and} \bibinfo{person}{Sandy~JJ
  Gould}.} \bibinfo{year}{2013}\natexlab{}.
\newblock \showarticletitle{Recovering from an interruption: Investigating
  speed- accuracy trade-offs in task resumption behavior.}
\newblock \bibinfo{journal}{\emph{Journal of Experimental Psychology: Applied}}
  \bibinfo{volume}{19}, \bibinfo{number}{2} (\bibinfo{year}{2013}),
  \bibinfo{pages}{95--107}.
\newblock


\bibitem[\protect\citeauthoryear{Colvin, Tobler, and Lindsay}{Colvin
  et~al\mbox{.}}{2004}]%
        {colvin2004productivity}
\bibfield{author}{\bibinfo{person}{Janet Colvin}, \bibinfo{person}{Nancy
  Tobler}, {and} \bibinfo{person}{Don Lindsay}.}
  \bibinfo{year}{2004}\natexlab{}.
\newblock \showarticletitle{Productivity and multi-screen displays}.
\newblock \bibinfo{journal}{\emph{Rocky Mountain Comm. Review 2: 1 2004}}
  (\bibinfo{year}{2004}), \bibinfo{pages}{31--53}.
\newblock


\bibitem[\protect\citeauthoryear{Czerwinski, Horvitz, and Wilhite}{Czerwinski
  et~al\mbox{.}}{2004}]%
        {czerwinski2004diary}
\bibfield{author}{\bibinfo{person}{Mary Czerwinski}, \bibinfo{person}{Eric
  Horvitz}, {and} \bibinfo{person}{Susan Wilhite}.}
  \bibinfo{year}{2004}\natexlab{}.
\newblock \showarticletitle{A Diary Study of Task Switching and Interruptions}.
  In \bibinfo{booktitle}{\emph{Proc. CHI}}. \bibinfo{publisher}{ACM},
  \bibinfo{pages}{175--182}.
\newblock
\showISBNx{1-58113-702-8}
\urldef\tempurl%
\url{https://doi.org/10.1145/985692.985715}
\showDOI{\tempurl}


\bibitem[\protect\citeauthoryear{Czerwinski, Smith, Regan, Meyers, Robertson,
  and Starkweather}{Czerwinski et~al\mbox{.}}{2003}]%
        {czerwinski2003toward}
\bibfield{author}{\bibinfo{person}{Mary Czerwinski}, \bibinfo{person}{Greg
  Smith}, \bibinfo{person}{Tim Regan}, \bibinfo{person}{Brian Meyers},
  \bibinfo{person}{George~G Robertson}, {and} \bibinfo{person}{Gary~K
  Starkweather}.} \bibinfo{year}{2003}\natexlab{}.
\newblock \showarticletitle{Toward characterizing the productivity benefits of
  very large displays.}. In \bibinfo{booktitle}{\emph{Interact}},
  Vol.~\bibinfo{volume}{3}. \bibinfo{pages}{9--16}.
\newblock


\bibitem[\protect\citeauthoryear{Duggan, Johnson, and S{\o}rli}{Duggan
  et~al\mbox{.}}{2013}]%
        {duggan2013interleaving}
\bibfield{author}{\bibinfo{person}{Geoffrey~B. Duggan}, \bibinfo{person}{Hilary
  Johnson}, {and} \bibinfo{person}{Petter S{\o}rli}.}
  \bibinfo{year}{2013}\natexlab{}.
\newblock \showarticletitle{Interleaving tasks to improve performance: Users
  maximise the marginal rate of return}.
\newblock \bibinfo{journal}{\emph{International. Journal of Human-Computer
  Studies}} \bibinfo{volume}{71}, \bibinfo{number}{5} (\bibinfo{year}{2013}),
  \bibinfo{pages}{533--550}.
\newblock
\showISSN{1071-5819}
\urldef\tempurl%
\url{https://doi.org/10.1016/j.ijhcs.2013.01.001}
\showDOI{\tempurl}


\bibitem[\protect\citeauthoryear{Fouse, Weibel, Hutchins, and Hollan}{Fouse
  et~al\mbox{.}}{2011}]%
        {fouse2011chronoviz}
\bibfield{author}{\bibinfo{person}{Adam Fouse}, \bibinfo{person}{Nadir Weibel},
  \bibinfo{person}{Edwin Hutchins}, {and} \bibinfo{person}{James~D. Hollan}.}
  \bibinfo{year}{2011}\natexlab{}.
\newblock \showarticletitle{ChronoViz: A System for Supporting Navigation of
  Time-coded Data}. In \bibinfo{booktitle}{\emph{CHI '11 Extended Abstracts on
  Human Factors in Computing Systems}} \emph{(\bibinfo{series}{CHI EA '11})}.
  \bibinfo{publisher}{ACM}, \bibinfo{address}{New York, NY, USA},
  \bibinfo{pages}{299--304}.
\newblock
\showISBNx{978-1-4503-0268-5}
\urldef\tempurl%
\url{https://doi.org/10.1145/1979742.1979706}
\showDOI{\tempurl}


\bibitem[\protect\citeauthoryear{Gonz\'{a}lez and Mark}{Gonz\'{a}lez and
  Mark}{2004}]%
        {gonzalez2004constant}
\bibfield{author}{\bibinfo{person}{Victor~M. Gonz\'{a}lez} {and}
  \bibinfo{person}{Gloria Mark}.} \bibinfo{year}{2004}\natexlab{}.
\newblock \showarticletitle{"Constant, Constant, Multi-tasking Craziness":
  Managing Multiple Working Spheres}. In \bibinfo{booktitle}{\emph{Proc. CHI}}.
  \bibinfo{publisher}{ACM}, \bibinfo{pages}{113--120}.
\newblock
\showISBNx{1-58113-702-8}
\urldef\tempurl%
\url{https://doi.org/10.1145/985692.985707}
\showDOI{\tempurl}


\bibitem[\protect\citeauthoryear{Gould, Brumby, and Cox}{Gould
  et~al\mbox{.}}{2013}]%
        {gould2013does}
\bibfield{author}{\bibinfo{person}{Sandy~JJ Gould}, \bibinfo{person}{Duncan~P
  Brumby}, {and} \bibinfo{person}{Anna~L Cox}.}
  \bibinfo{year}{2013}\natexlab{}.
\newblock \showarticletitle{What does it mean for an interruption to be
  relevant? An investigation of relevance as a memory effect}. In
  \bibinfo{booktitle}{\emph{Proceedings of the Human Factors and Ergonomics
  Society Annual Meeting}}, Vol.~\bibinfo{volume}{57}. SAGE Publications,
  \bibinfo{pages}{149--153}.
\newblock


\bibitem[\protect\citeauthoryear{Grudin}{Grudin}{2001}]%
        {grudin2001partitioning}
\bibfield{author}{\bibinfo{person}{Jonathan Grudin}.}
  \bibinfo{year}{2001}\natexlab{}.
\newblock \showarticletitle{Partitioning Digital Worlds: Focal and Peripheral
  Awareness in Multiple Monitor Use}. In \bibinfo{booktitle}{\emph{Proceedings
  of the SIGCHI Conference on Human Factors in Computing Systems}}
  \emph{(\bibinfo{series}{CHI '01})}. \bibinfo{publisher}{ACM},
  \bibinfo{address}{New York, NY, USA}, \bibinfo{pages}{458--465}.
\newblock
\showISBNx{1-58113-327-8}
\urldef\tempurl%
\url{https://doi.org/10.1145/365024.365312}
\showDOI{\tempurl}


\bibitem[\protect\citeauthoryear{Hardy and Gillan}{Hardy and Gillan}{2012}]%
        {hardy2012voluntary}
\bibfield{author}{\bibinfo{person}{Megan Hardy} {and}
  \bibinfo{person}{Douglas~J Gillan}.} \bibinfo{year}{2012}\natexlab{}.
\newblock \showarticletitle{Voluntary task switching patterns in everyday tasks
  of different motivational levels}. In \bibinfo{booktitle}{\emph{Proceedings
  of the Human Factors and Ergonomics Society Annual Meeting}},
  Vol.~\bibinfo{volume}{56}. SAGE Publications, \bibinfo{pages}{2128--2132}.
\newblock


\bibitem[\protect\citeauthoryear{Henderson and Card}{Henderson and
  Card}{1986}]%
        {henderson1986rooms}
\bibfield{author}{\bibinfo{person}{D.~Austin Henderson, Jr.} {and}
  \bibinfo{person}{Stuart Card}.} \bibinfo{year}{1986}\natexlab{}.
\newblock \showarticletitle{Rooms: The Use of Multiple Virtual Workspaces to
  Reduce Space Contention in a Window-based Graphical User Interface}.
\newblock \bibinfo{journal}{\emph{ACM Trans. Graph.}} \bibinfo{volume}{5},
  \bibinfo{number}{3} (\bibinfo{date}{July} \bibinfo{year}{1986}),
  \bibinfo{pages}{211--243}.
\newblock
\showISSN{0730-0301}
\urldef\tempurl%
\url{https://doi.org/10.1145/24054.24056}
\showDOI{\tempurl}


\bibitem[\protect\citeauthoryear{Houben, Bardram, Vermeulen, Luyten, and
  Coninx}{Houben et~al\mbox{.}}{2013}]%
        {houben2013cam}
\bibfield{author}{\bibinfo{person}{Steven Houben}, \bibinfo{person}{Jakob~E.
  Bardram}, \bibinfo{person}{Jo Vermeulen}, \bibinfo{person}{Kris Luyten},
  {and} \bibinfo{person}{Karin Coninx}.} \bibinfo{year}{2013}\natexlab{}.
\newblock \showarticletitle{Activity-centric Support for Ad Hoc Knowledge Work:
  A Case Study of Co-activity Manager}. In \bibinfo{booktitle}{\emph{Proc.
  CHI}}. \bibinfo{publisher}{ACM}, \bibinfo{pages}{2263--2272}.
\newblock
\showISBNx{978-1-4503-1899-0}
\urldef\tempurl%
\url{https://doi.org/10.1145/2470654.2481312}
\showDOI{\tempurl}


\bibitem[\protect\citeauthoryear{Hutchings, Smith, Meyers, Czerwinski, and
  Robertson}{Hutchings et~al\mbox{.}}{2004}]%
        {hutchings2004displayspace}
\bibfield{author}{\bibinfo{person}{Dugald~Ralph Hutchings},
  \bibinfo{person}{Greg Smith}, \bibinfo{person}{Brian Meyers},
  \bibinfo{person}{Mary Czerwinski}, {and} \bibinfo{person}{George Robertson}.}
  \bibinfo{year}{2004}\natexlab{}.
\newblock \showarticletitle{Display Space Usage and Window Management Operation
  Comparisons Between Single Monitor and Multiple Monitor Users}. In
  \bibinfo{booktitle}{\emph{Proceedings of the Working Conference on Advanced
  Visual Interfaces}} \emph{(\bibinfo{series}{AVI '04})}.
  \bibinfo{publisher}{ACM}, \bibinfo{address}{New York, NY, USA},
  \bibinfo{pages}{32--39}.
\newblock
\showISBNx{1-58113-867-9}
\urldef\tempurl%
\url{https://doi.org/10.1145/989863.989867}
\showDOI{\tempurl}


\bibitem[\protect\citeauthoryear{Hutchings and Stasko}{Hutchings and
  Stasko}{2004}]%
        {hutchings2004revisiting}
\bibfield{author}{\bibinfo{person}{Dugald~Ralph Hutchings} {and}
  \bibinfo{person}{John Stasko}.} \bibinfo{year}{2004}\natexlab{}.
\newblock \showarticletitle{Revisiting display space management: understanding
  current practice to inform next-generation design}. In
  \bibinfo{booktitle}{\emph{Proceedings of Graphics interface 2004}}. Canadian
  Human-Computer Communications Society, \bibinfo{pages}{127--134}.
\newblock


\bibitem[\protect\citeauthoryear{Iqbal and Horvitz}{Iqbal and Horvitz}{2007}]%
        {iqbal2007disruption}
\bibfield{author}{\bibinfo{person}{Shamsi~T. Iqbal} {and} \bibinfo{person}{Eric
  Horvitz}.} \bibinfo{year}{2007}\natexlab{}.
\newblock \showarticletitle{Disruption and Recovery of Computing Tasks: Field
  Study, Analysis, and Directions}. In \bibinfo{booktitle}{\emph{Proceedings of
  the SIGCHI Conference on Human Factors in Computing Systems}}
  \emph{(\bibinfo{series}{CHI '07})}. \bibinfo{publisher}{ACM},
  \bibinfo{address}{New York, NY, USA}, \bibinfo{pages}{677--686}.
\newblock
\showISBNx{978-1-59593-593-9}
\urldef\tempurl%
\url{https://doi.org/10.1145/1240624.1240730}
\showDOI{\tempurl}


\bibitem[\protect\citeauthoryear{Janssen, Gould, Li, Brumby, and Cox}{Janssen
  et~al\mbox{.}}{2015}]%
        {janssen2015integrating}
\bibfield{author}{\bibinfo{person}{Christian~P Janssen},
  \bibinfo{person}{Sandy~JJ Gould}, \bibinfo{person}{Simon~YW Li},
  \bibinfo{person}{Duncan~P Brumby}, {and} \bibinfo{person}{Anna~L Cox}.}
  \bibinfo{year}{2015}\natexlab{}.
\newblock \showarticletitle{Integrating knowledge of multitasking and
  Interruptions across different Perspectives and research methods}.
\newblock \bibinfo{journal}{\emph{International Journal of Human-Computer
  Studies}}  \bibinfo{volume}{79} (\bibinfo{year}{2015}),
  \bibinfo{pages}{1--5}.
\newblock


\bibitem[\protect\citeauthoryear{Jeuris and Bardram}{Jeuris and
  Bardram}{2016}]%
        {jeuris2016dedicated}
\bibfield{author}{\bibinfo{person}{Steven Jeuris} {and}
  \bibinfo{person}{Jakob~E. Bardram}.} \bibinfo{year}{2016}\natexlab{}.
\newblock \showarticletitle{Dedicated workspaces: Faster resumption times and
  reduced cognitive load in sequential multitasking}.
\newblock \bibinfo{journal}{\emph{Computers in Human Behavior}}
  \bibinfo{volume}{62} (\bibinfo{year}{2016}), \bibinfo{pages}{404--414}.
\newblock
\showISSN{0747-5632}
\urldef\tempurl%
\url{https://doi.org/10.1016/j.chb.2016.03.059}
\showDOI{\tempurl}


\bibitem[\protect\citeauthoryear{Jeuris, Houben, and Bardram}{Jeuris
  et~al\mbox{.}}{2014}]%
        {jeuris2014laevo}
\bibfield{author}{\bibinfo{person}{Steven Jeuris}, \bibinfo{person}{Steven
  Houben}, {and} \bibinfo{person}{Jakob~E. Bardram}.}
  \bibinfo{year}{2014}\natexlab{}.
\newblock \showarticletitle{Laevo: A Temporal Desktop Interface for Integrated
  Knowledge Work}. In \bibinfo{booktitle}{\emph{Proc. UIST}}. ACM.
\newblock


\bibitem[\protect\citeauthoryear{Jin and Dabbish}{Jin and Dabbish}{2009}]%
        {jin2009selfinterruption}
\bibfield{author}{\bibinfo{person}{Jing Jin} {and} \bibinfo{person}{Laura~A.
  Dabbish}.} \bibinfo{year}{2009}\natexlab{}.
\newblock \showarticletitle{Self-interruption on the Computer: A Typology of
  Discretionary Task Interleaving}. In \bibinfo{booktitle}{\emph{Proceedings of
  the SIGCHI Conference on Human Factors in Computing Systems}}
  \emph{(\bibinfo{series}{CHI '09})}. \bibinfo{publisher}{ACM},
  \bibinfo{pages}{1799--1808}.
\newblock
\showISBNx{978-1-60558-246-7}
\urldef\tempurl%
\url{https://doi.org/10.1145/1518701.1518979}
\showDOI{\tempurl}


\bibitem[\protect\citeauthoryear{Kang and Stasko}{Kang and Stasko}{2008}]%
        {kang2008lightweight}
\bibfield{author}{\bibinfo{person}{Youn-ah Kang} {and} \bibinfo{person}{John
  Stasko}.} \bibinfo{year}{2008}\natexlab{}.
\newblock \showarticletitle{Lightweight task/application performance using
  single versus multiple monitors: a comparative study}. In
  \bibinfo{booktitle}{\emph{Proceedings of Graphics Interface 2008}}. Canadian
  Information Processing Society, \bibinfo{pages}{17--24}.
\newblock


\bibitem[\protect\citeauthoryear{Ling, Stegman, Barhbaya, and Shehab}{Ling
  et~al\mbox{.}}{2017}]%
        {ling2017twobetter}
\bibfield{author}{\bibinfo{person}{Chen Ling}, \bibinfo{person}{Alex Stegman},
  \bibinfo{person}{Chintan Barhbaya}, {and} \bibinfo{person}{Randa Shehab}.}
  \bibinfo{year}{2017}\natexlab{}.
\newblock \showarticletitle{Are Two Better Than One? A Comparison Between
  Single- and Dual-Monitor Work Stations in Productivity and User’s Windows
  Management Style}.
\newblock \bibinfo{journal}{\emph{International Journal of Human–Computer
  Interaction}} \bibinfo{volume}{33}, \bibinfo{number}{3}
  (\bibinfo{year}{2017}), \bibinfo{pages}{190--198}.
\newblock
\urldef\tempurl%
\url{https://doi.org/10.1080/10447318.2016.1231392}
\showDOI{\tempurl}
\showeprint{http://dx.doi.org/10.1080/10447318.2016.1231392}


\bibitem[\protect\citeauthoryear{Marien, Custers, Hassin, and Aarts}{Marien
  et~al\mbox{.}}{2012}]%
        {marien2012unconscious}
\bibfield{author}{\bibinfo{person}{Hans Marien}, \bibinfo{person}{Ruud
  Custers}, \bibinfo{person}{Ran~R Hassin}, {and} \bibinfo{person}{Henk
  Aarts}.} \bibinfo{year}{2012}\natexlab{}.
\newblock \showarticletitle{Unconscious goal activation and the hijacking of
  the executive function.}
\newblock \bibinfo{journal}{\emph{Journal of personality and social
  psychology}} \bibinfo{volume}{103}, \bibinfo{number}{3}
  (\bibinfo{year}{2012}), \bibinfo{pages}{399--415}.
\newblock


\bibitem[\protect\citeauthoryear{Mark, Gudith, and Klocke}{Mark
  et~al\mbox{.}}{2008}]%
        {mark2008stress}
\bibfield{author}{\bibinfo{person}{Gloria Mark}, \bibinfo{person}{Daniela
  Gudith}, {and} \bibinfo{person}{Ulrich Klocke}.}
  \bibinfo{year}{2008}\natexlab{}.
\newblock \showarticletitle{The Cost of Interrupted Work: More Speed and
  Stress}. In \bibinfo{booktitle}{\emph{Proceedings of the SIGCHI Conference on
  Human Factors in Computing Systems}} \emph{(\bibinfo{series}{CHI '08})}.
  \bibinfo{publisher}{ACM}, \bibinfo{pages}{107--110}.
\newblock
\showISBNx{978-1-60558-011-1}
\urldef\tempurl%
\url{https://doi.org/10.1145/1357054.1357072}
\showDOI{\tempurl}


\bibitem[\protect\citeauthoryear{Mark, Voida, and Cardello}{Mark
  et~al\mbox{.}}{2012}]%
        {mark2012email}
\bibfield{author}{\bibinfo{person}{Gloria Mark}, \bibinfo{person}{Stephen
  Voida}, {and} \bibinfo{person}{Armand Cardello}.}
  \bibinfo{year}{2012}\natexlab{}.
\newblock \showarticletitle{"{A} Pace Not Dictated by Electrons": An Empirical
  Study of Work Without Email}. In \bibinfo{booktitle}{\emph{Proceedings of the
  SIGCHI Conference on Human Factors in Computing Systems}}
  \emph{(\bibinfo{series}{CHI '12})}. \bibinfo{publisher}{ACM},
  \bibinfo{pages}{555--564}.
\newblock
\showISBNx{978-1-4503-1015-4}
\urldef\tempurl%
\url{https://doi.org/10.1145/2207676.2207754}
\showDOI{\tempurl}


\bibitem[\protect\citeauthoryear{Masicampo and Baumeister}{Masicampo and
  Baumeister}{2011a}]%
        {masicampo2011consider}
\bibfield{author}{\bibinfo{person}{EJ Masicampo} {and} \bibinfo{person}{Roy~F
  Baumeister}.} \bibinfo{year}{2011}\natexlab{a}.
\newblock \showarticletitle{Consider it done! Plan making can eliminate the
  cognitive effects of unfulfilled goals.}
\newblock \bibinfo{journal}{\emph{Journal of personality and social
  psychology}} \bibinfo{volume}{101}, \bibinfo{number}{4}
  (\bibinfo{year}{2011}), \bibinfo{pages}{667--683}.
\newblock


\bibitem[\protect\citeauthoryear{Masicampo and Baumeister}{Masicampo and
  Baumeister}{2011b}]%
        {masicampo2011unfulfilled}
\bibfield{author}{\bibinfo{person}{EJ Masicampo} {and} \bibinfo{person}{Roy~F
  Baumeister}.} \bibinfo{year}{2011}\natexlab{b}.
\newblock \showarticletitle{Unfulfilled goals interfere with tasks that require
  executive functions}.
\newblock \bibinfo{journal}{\emph{Journal of Experimental Social Psychology}}
  \bibinfo{volume}{47}, \bibinfo{number}{2} (\bibinfo{year}{2011}),
  \bibinfo{pages}{300--311}.
\newblock


\bibitem[\protect\citeauthoryear{Monk, Trafton, and Boehm-Davis}{Monk
  et~al\mbox{.}}{2008}]%
        {monk2008effect}
\bibfield{author}{\bibinfo{person}{Christopher~A Monk},
  \bibinfo{person}{J~Gregory Trafton}, {and} \bibinfo{person}{Deborah~A
  Boehm-Davis}.} \bibinfo{year}{2008}\natexlab{}.
\newblock \showarticletitle{The effect of interruption duration and demand on
  resuming suspended goals.}
\newblock \bibinfo{journal}{\emph{Journal of Experimental Psychology: Applied}}
  \bibinfo{volume}{14}, \bibinfo{number}{4} (\bibinfo{year}{2008}),
  \bibinfo{pages}{299--313}.
\newblock


\bibitem[\protect\citeauthoryear{Mulder, de~Poot, Verwij, Janssen, and
  Bijlsma}{Mulder et~al\mbox{.}}{2006}]%
        {mulder2006information}
\bibfield{author}{\bibinfo{person}{Ingrid Mulder}, \bibinfo{person}{Henk de
  Poot}, \bibinfo{person}{Carla Verwij}, \bibinfo{person}{Ruud Janssen}, {and}
  \bibinfo{person}{Marcel Bijlsma}.} \bibinfo{year}{2006}\natexlab{}.
\newblock \showarticletitle{An information overload study: using design methods
  for understanding}. In \bibinfo{booktitle}{\emph{Proc. OZCHI}}. ACM,
  \bibinfo{pages}{245--252}.
\newblock


\bibitem[\protect\citeauthoryear{Owens, Teves, Nguyen, Smith, Phelps, and
  Chaparro}{Owens et~al\mbox{.}}{2012}]%
        {owens2012examination}
\bibfield{author}{\bibinfo{person}{Justin~W. Owens}, \bibinfo{person}{Jennifer
  Teves}, \bibinfo{person}{Bobby Nguyen}, \bibinfo{person}{Amanda Smith},
  \bibinfo{person}{Mandy~C. Phelps}, {and} \bibinfo{person}{Barbara~S.
  Chaparro}.} \bibinfo{year}{2012}\natexlab{}.
\newblock \showarticletitle{Examination of Dual vs. Single Monitor Use during
  Common Office Tasks}.
\newblock \bibinfo{journal}{\emph{Proceedings of the Human Factors and
  Ergonomics Society Annual Meeting}} \bibinfo{volume}{56}, \bibinfo{number}{1}
  (\bibinfo{year}{2012}), \bibinfo{pages}{1506--1510}.
\newblock
\urldef\tempurl%
\url{https://doi.org/10.1177/1071181312561299}
\showDOI{\tempurl}
\showeprint{https://doi.org/10.1177/1071181312561299}


\bibitem[\protect\citeauthoryear{Payne, Duggan, and Neth}{Payne
  et~al\mbox{.}}{2007}]%
        {payne2007discretionary}
\bibfield{author}{\bibinfo{person}{Stephen~J Payne},
  \bibinfo{person}{Geoffrey~B Duggan}, {and} \bibinfo{person}{Hansj{\"o}rg
  Neth}.} \bibinfo{year}{2007}\natexlab{}.
\newblock \showarticletitle{Discretionary task interleaving: heuristics for
  time allocation in cognitive foraging.}
\newblock \bibinfo{journal}{\emph{Journal of Experimental Psychology: General}}
  \bibinfo{volume}{136}, \bibinfo{number}{3} (\bibinfo{year}{2007}),
  \bibinfo{pages}{370--388}.
\newblock
\urldef\tempurl%
\url{https://doi.org/10.1037/0096-3445.136.3.370}
\showDOI{\tempurl}


\bibitem[\protect\citeauthoryear{Ringel}{Ringel}{2003}]%
        {ringel2003analysis}
\bibfield{author}{\bibinfo{person}{Meredith Ringel}.}
  \bibinfo{year}{2003}\natexlab{}.
\newblock \showarticletitle{When One Isn't Enough: An Analysis of Virtual
  Desktop Usage Strategies and Their Implications for Design}. In
  \bibinfo{booktitle}{\emph{CHI EA}}. \bibinfo{publisher}{ACM},
  \bibinfo{pages}{762--763}.
\newblock
\showISBNx{1-58113-637-4}
\urldef\tempurl%
\url{https://doi.org/10.1145/765891.765976}
\showDOI{\tempurl}


\bibitem[\protect\citeauthoryear{Robertson, Horvitz, Czerwinski, Baudisch,
  Hutchings, Meyers, Robbins, and Smith}{Robertson et~al\mbox{.}}{2004}]%
        {robertson2004scalable}
\bibfield{author}{\bibinfo{person}{George Robertson}, \bibinfo{person}{Eric
  Horvitz}, \bibinfo{person}{Mary Czerwinski}, \bibinfo{person}{Patrick
  Baudisch}, \bibinfo{person}{Dugald~Ralph Hutchings}, \bibinfo{person}{Brian
  Meyers}, \bibinfo{person}{Daniel Robbins}, {and} \bibinfo{person}{Greg
  Smith}.} \bibinfo{year}{2004}\natexlab{}.
\newblock \showarticletitle{Scalable Fabric: Flexible Task Management}. In
  \bibinfo{booktitle}{\emph{Proc. AVI}}. \bibinfo{publisher}{ACM},
  \bibinfo{pages}{85--89}.
\newblock
\showISBNx{1-58113-867-9}
\urldef\tempurl%
\url{https://doi.org/10.1145/989863.989874}
\showDOI{\tempurl}


\bibitem[\protect\citeauthoryear{Salvucci, Taatgen, and Borst}{Salvucci
  et~al\mbox{.}}{2009}]%
        {salvucci2009towards}
\bibfield{author}{\bibinfo{person}{Dario~D. Salvucci},
  \bibinfo{person}{Niels~A. Taatgen}, {and} \bibinfo{person}{Jelmer~P. Borst}.}
  \bibinfo{year}{2009}\natexlab{}.
\newblock \showarticletitle{Toward a Unified Theory of the Multitasking
  Continuum: From Concurrent Performance to Task Switching, Interruption, and
  Resumption}. In \bibinfo{booktitle}{\emph{Proc. CHI}}.
  \bibinfo{publisher}{ACM}, \bibinfo{pages}{1819--1828}.
\newblock
\showISBNx{978-1-60558-246-7}
\urldef\tempurl%
\url{https://doi.org/10.1145/1518701.1518981}
\showDOI{\tempurl}


\bibitem[\protect\citeauthoryear{Strayer and Johnston}{Strayer and
  Johnston}{2001}]%
        {strayer2001driven}
\bibfield{author}{\bibinfo{person}{David~L Strayer} {and}
  \bibinfo{person}{William~A Johnston}.} \bibinfo{year}{2001}\natexlab{}.
\newblock \showarticletitle{Driven to distraction: Dual-task studies of
  simulated driving and conversing on a cellular telephone}.
\newblock \bibinfo{journal}{\emph{Psychological science}} \bibinfo{volume}{12},
  \bibinfo{number}{6} (\bibinfo{year}{2001}), \bibinfo{pages}{462--466}.
\newblock


\bibitem[\protect\citeauthoryear{Tashman and Edwards}{Tashman and
  Edwards}{2012}]%
        {tashman2012windowscape}
\bibfield{author}{\bibinfo{person}{Craig Tashman} {and}
  \bibinfo{person}{W.~Keith Edwards}.} \bibinfo{year}{2012}\natexlab{}.
\newblock \showarticletitle{WindowScape: Lessons Learned from a Task-centric
  Window Manager}.
\newblock \bibinfo{journal}{\emph{ACM Transactions on Computer-Human
  Interaction}} \bibinfo{volume}{19}, \bibinfo{number}{1}, Article
  \bibinfo{articleno}{8} (\bibinfo{date}{May} \bibinfo{year}{2012}),
  \bibinfo{numpages}{33}~pages.
\newblock
\showISSN{1073-0516}
\urldef\tempurl%
\url{https://doi.org/10.1145/2147783.2147791}
\showDOI{\tempurl}


\bibitem[\protect\citeauthoryear{Trafton, Altmann, Brock, and Mintz}{Trafton
  et~al\mbox{.}}{2003}]%
        {trafton2003preparing}
\bibfield{author}{\bibinfo{person}{J.Gregory Trafton}, \bibinfo{person}{Erik~M
  Altmann}, \bibinfo{person}{Derek~P Brock}, {and} \bibinfo{person}{Farilee~E
  Mintz}.} \bibinfo{year}{2003}\natexlab{}.
\newblock \showarticletitle{Preparing to resume an interrupted task: effects of
  prospective goal encoding and retrospective rehearsal}.
\newblock \bibinfo{journal}{\emph{International Journal of Human-Computer
  Studies}} \bibinfo{volume}{58}, \bibinfo{number}{5} (\bibinfo{year}{2003}),
  \bibinfo{pages}{583--603}.
\newblock
\showISSN{1071-5819}
\urldef\tempurl%
\url{https://doi.org/10.1016/S1071-5819(03)00023-5}
\showDOI{\tempurl}


\bibitem[\protect\citeauthoryear{Voida, Mynatt, and Edwards}{Voida
  et~al\mbox{.}}{2008}]%
        {voida2008reframing}
\bibfield{author}{\bibinfo{person}{Stephen Voida},
  \bibinfo{person}{Elizabeth~D. Mynatt}, {and} \bibinfo{person}{W.~Keith
  Edwards}.} \bibinfo{year}{2008}\natexlab{}.
\newblock \showarticletitle{Re-framing the Desktop Interface Around the
  Activities of Knowledge Work}. In \bibinfo{booktitle}{\emph{Proc. UIST}}.
  \bibinfo{publisher}{ACM}, \bibinfo{pages}{211--220}.
\newblock
\showISBNx{978-1-59593-975-3}
\urldef\tempurl%
\url{https://doi.org/10.1145/1449715.1449751}
\showDOI{\tempurl}


\end{thebibliography}

\end{document}